\documentclass[sigconf]{acmart}
\AtBeginDocument{%
}

\copyrightyear{2024}
\acmYear{2024}
\setcopyright{rightsretained}
\acmConference[IMC '24]{Proceedings of the 2024 ACM Internet Measurement Conference}{November 4--6, 2024}{Madrid, Spain}
\acmBooktitle{Proceedings of the 2024 ACM Internet Measurement Conference (IMC '24), November 4--6, 2024, Madrid, Spain}\acmDOI{10.1145/3646547.3688452}
\acmISBN{979-8-4007-0592-2/24/11}

\makeatletter
\gdef\@copyrightpermission{
 \begin{minipage}{0.3\columnwidth}
 \href{https://creativecommons.org/licenses/by/4.0/}{\includegraphics[width=0.90\textwidth]{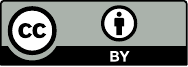}}
 \end{minipage}\hfill
 \begin{minipage}{0.7\columnwidth}
 \href{https://creativecommons.org/licenses/by/4.0/}{This work is licensed under a Creative Commons
Attribution International 4.0 License.}
 \end{minipage}
 \vspace{5pt}
}
\makeatother

\usepackage[english]{babel}
\usepackage{blindtext}
\usepackage{subcaption}
\usepackage{lscape}
\usepackage{booktabs}
\usepackage{pifont}
\usepackage{makecell}
\usepackage{cleveref}
\usepackage{siunitx}
\crefformat{section}{\S#2#1#3}
\usepackage{enumitem}
\usepackage{multirow}
\usepackage{balance}

\newif\ifshowcomments
\showcommentsfalse

\newcommand{\rev}[1]{\ifshowcomments {\leavevmode\color{blue}#1} \else {#1}\fi\unskip}%
\newcommand{\revtwo}[1]{\ifshowcomments {\leavevmode\color{blue}#1} \else {#1}\fi\unskip}%
\newcommand{\revthree}[1]{\ifshowcomments {\leavevmode\color{red}#1} \else {#1}\fi\unskip}%
\newif\ifcomment
\commenttrue % uncomment this to enable comments
\ifcomment
    \newcounter{aelNumberOfComments}
    \stepcounter{aelNumberOfComments}

\fi

\begin{document}

\title[Through the Telco Lens: A Countrywide Empirical Study of Cellular Handovers]{Through the Telco Lens: A Countrywide Empirical Study of Cellular Handovers}

\author{Michail Kalntis}
\affiliation{%
\institution{Delft University of Technology}
\city{Delft}
\country{the Netherlands}
}
\email{m.kalntis@tudelft.nl}

\author{Jos\'e Su\'arez-Varela}
\affiliation{%
\institution{Telef\'onica Research}
\city{Madrid}
\country{Spain}
}
\email{jose.suarez-varela@telefonica.com}

\author{Jes\'us Oma\~na Iglesias}
\affiliation{%
\institution{Telef\'onica Research}
\city{Barcelona}
\country{Spain}
}
\email{jesusalberto.omana@telefonica.com}

\author{Anup Kiran Bhattacharjee}
\affiliation{%
\institution{Delft University of Technology}
\city{Delft}
\country{the Netherlands}
}
\email{a.k.bhattacharjee@tudelft.nl}

\author{George Iosifidis}
\affiliation{%
\institution{Delft University of Technology}
\city{Delft}
\country{the Netherlands}
}
\email{g.iosifidis@tudelft.nl}

\author{Fernando A. Kuipers}
\affiliation{%
\institution{Delft University of Technology}
\city{Delft}
\country{the Netherlands}
}
\email{f.a.kuipers@tudelft.nl}

\author{Andra Lutu}
\affiliation{%
\institution{Telef\'onica Research}
\city{Madrid}
\country{Spain}
}
\email{andra.lutu@telefonica.com}

\renewcommand{\shortauthors}{Michail Kalntis et al.}

\begin{abstract}
Cellular networks rely on handovers (HOs) as a fundamental element to enable seamless connectivity for mobile users. A comprehensive analysis of HOs can be achieved through data from Mobile Network Operators (MNOs); however, the vast majority of studies employ data from measurement campaigns within confined areas and with limited end-user devices, thereby providing only a partial view of HOs. This paper presents the first countrywide analysis of HO performance, from the perspective of a top-tier MNO in a European country. We collect traffic from approximately 40M users for 4 weeks and study the impact of the radio access technologies (RATs), device types, and manufacturers on HOs across the country. We characterize the geo-temporal dynamics of horizontal (intra-RAT) and vertical (inter-RATs) HOs, at the district level and at millisecond granularity, and leverage open datasets from the country's official census office to associate our findings with the population. We further delve into the frequency, duration, and causes of HO failures, and model them using statistical tools. Our study offers unique insights into mobility management, highlighting the heterogeneity of the network and devices, and their effect on HOs.

\end{abstract}




\maketitle

\section{Introduction}\label{sec:introduction}

The advent of the 5G New Radio (NR) technology marked a shift in the telecommunications landscape, offering unprecedented speed and widespread connectivity to a vast array of devices~\cite{5g-ngmn}. With these developments, end-user expectations have surged, driven by the promise of higher bandwidth, lower latency, and importantly, ubiquitous connectivity for fast-moving User Equipment (UE) through effective handovers (HOs). However, like any emerging technology, the pace of real-world deployments does not instantly match the pace of innovation~\cite{polese2023empowering, mahimkar2021auric}, resulting in multiple generations of technology operated simultaneously to balance the trade-offs between OPEX/CAPEX and the stringent needs for high availability, reliability, and capacity. In this complex arena, HOs have become more intricate yet crucial for maintaining seamless connectivity.

Prior works study the intricacies behind HO management~\cite{li2020beyond, china_20, vivisecting_2022}, yet the majority of them \cite{mob_support_2018, lte_rails_2019, china_20, raca_4G_dataset, xiao_4G_dataset, li2020beyond, 5G_ues_2020, 5G_wild_2021, 5G_mmwave_2022, raca_5G_dataset, vivisecting_2022, wheels_2023, li2021nationwide} analyze HOs based on limited user-side measurement campaigns and, as such, are confined to specific mobility scenarios, geographic areas, or UE manufacturers. These limitations restrict the conclusions we can draw and underscore the need for a detailed, large-scale analysis that captures the complexity and heterogeneity behind real MNO deployments.

The goal of this paper is to fill the existing gap by presenting the \textit{first, to our knowledge, countrywide analysis of mobility management, from the perspective of a top-tier Mobile Network Operator (MNO) \revtwo{in Europe}}.\footnote{\revtwo{To maintain the confidentiality of the operator, we are only able to disclose general location information for the study.}} We have recorded \emph{all mobility events} over a period of 4 weeks at millisecond granularity, across the entire country. Our datasets (see Table \ref{tab:statistics}) include all HOs and HO failures (HOFs) that occurred over the observed period, enabling the analysis of network dynamics at a crucial moment when -- at the time of writing -- \textit{all} digital radio access technologies (RATs) developed during the last three decades are concurrently operational within the same network. We merge this data with: ($i$)~information from the MNO's deployment, to study HO performance across its topology and supported RATs, ($ii$)~device-specific information, to associate HOs and HOFs with specific UE types and manufacturers, and ($iii$)~the data from the \revtwo{country's official census office,} to account for the geodemographic distribution of HOs across \revtwo{300+ districts with various population density.} At the time of the study, the MNO was just initiating its commercial 5G-Standalone (SA) deployment, so we measured only the 5G-Non-Standalone (NSA) deployment to avoid any early-stage issues with SA~\cite{SA_vs_NSA}.

\begin{table}[!t]
{
\small{
    \centering
    \caption{\revtwo{Dataset statistics.}}
    \vspace{-2mm}
    \begin{tabular}{p{3cm}p{4.5cm}}
    
    \toprule
    Feature & Value \\
    
    \midrule
    Area covered & \revtwo{Country in Europe (300+ districts)} \\
    \# of cell sites & $24$\small{k}\footnotesize{$+$} \\
    \# of radio sectors & $350$\small{k}\footnotesize{$+$} \\
    \# of UEs measured & $\approx$$40$\small{M} \\
    \# handovers (daily) & $1.7$\small{B}\footnotesize{$+$} \\
    Measurement duration & 4 weeks (28 days) \\
    Trace size (daily) & $\approx$\SI{8}{\tera\byte} \\
    
    \bottomrule\bottomrule
    \end{tabular}
    \label{tab:statistics}
}
}
\end{table}

To capture the complexity and heterogeneity behind the studied network-side datasets, we define three main dimensions that significantly affect HOs and mobility management: ($i$)~the heterogeneity of RATs in the MNO, namely 2G, 3G, 4G, and 5G,\footnote{The Second, Third, Fourth, and Fifth Generation networks, and their respective RATs, are henceforth referred as 2G, 3G, 4G, and 5G, respectively.} ($ii$)~the heterogeneity of UEs, and ($iii$)~the geodemographic diversity. We analyze the spatio-temporal dynamics of horizontal (intra-RAT), and vertical (inter-RAT) HOs, at the district-level and with msec granularity, and characterize their pattern across the country to identify regional trends. Furthermore, we dissect the impact of UE types (smartphones, M2M/IoT\footnote{See Appendix~\ref{sec:appendix-abbreviations} for the meaning of all abbreviations (e.g., M2M/IoT).} devices, low-tier feature phones) and manufacturers on HOs, HOFs, and mobility/performance metrics. We also analyze the causes behind HOFs, using 3GPP-based and vendor-specific failure descriptions. Lastly, we leverage statistical methods to model how the coexistence of multiple RATs affects HO performance, especially when UE connections are downgraded to older technologies (e.g., 2G, 3G). Below we present the key findings and contributions
of this work.

\noindent$\bullet$ \textbf{Heterogeneity \& Complexity of HOs (\S\ref{sec:first_look_at_network})}. 
In the MNO's deployment, 5G sectors make up 8.4\% while 4G accounts for 55\%, with 2G and 3G sectors covering the remaining $\approx$36\% and handling 18\% of user connectivity time. Despite this, older RATs carry only 5.23\% of the uplink (UL) and 2.07\% of the downlink (DL) data flowing through the network. Among all UEs, 59.1\% are smartphones, primarily from Apple (54.8\%) and Samsung (30.2\%), from which 51.5\% do not support 5G, relying instead on 4G. Additionally, over 32\% of the UEs, mainly M2M/IoT devices and feature phones, support only up to 3G. This blend of technologies highlights the challenges of phasing out older RATs, particularly in an environment where IoT manufacturers still rely on 3G/2G for devices with limited connectivity needs. Our geodemographic analysis points to a large disparity between the density of HOs in urban centers \revtwo{with larger population density} (2.1M HOs per sq. km), and \revtwo{less populated} rural areas (60 HOs per sq. km); in a network deployment that registers on average 13.1k HOs per sq. km.

\noindent$\bullet$ \textbf{HO Analysis (\S\ref{sec:mobility_analysis})}. 
Taking as a reference the HOs registered in the 4G EPC, approximately 94\% of HOs are horizontal (between 4G/5G-NSA radio sectors), complete within 90 ms (median of 43ms), and correspond to smartphone activity. M2M/IoT UEs and low-tier feature phones --~accounting for $>\!\!40\%$ of the device population~-- share the remaining 6\% of HOs. By investigating the top-5 smartphone manufacturers (Apple, Samsung, Motorola, Google, Huawei), we discover similar patterns in terms of HOs ($\pm$10\% of variation between them) and low HOF rates (Google exhibits -27\% of HOFs w.r.t. other UEs, but with higher variability). Moreover, we find some smartphone manufacturers outside the top-5 (e.g., KVD) that exhibit higher HOF rates (up to +600\% w.r.t. other UEs) and HO signaling (up to +293\%).

\noindent$\bullet$ \textbf{HOF Analysis \& Modeling (\S\ref{sec:areas_of_attention})}.
Rural areas (with sparser deployments) experience 32.4\% more HOFs during peak hours [7:00--8:00) than urban areas. Moreover, HOF rate is close to zero for the majority of the UEs; for the ones with high mobility metrics ($>$100 visited sectors, $>$100km radius of gyration), which are mostly smartphones (85\%), HOF rate rises up to 0.4\% (pct-75).

Furthermore, we dissect the reasons why HOs fail by using 1k+ 3GPP and vendor-specific descriptions that explain the causes. Interestingly, we find that 92\% of the HOs in the entire country fail with solely 8 causes; and from the studied failures, 75\% (0.03\%) occur in HOs to 3G (2G), and 25\% of them are due to an excessive load in the target sector (Cause~\#4). Moreover, we measure the duration of these 8 causes and highlight that the ones related to specific cancellations (Cause~\#1) and timeouts (Cause~\#8) require on average >2s to complete, reaching up to 10s in the latter case.

On top of the previous analysis, we aim to unveil what network-related features correlate with HOFs. Specifically, we test whether the HO type (intra/inter-RAT) is a good predictor for the HOF rate. Statistical analysis verifies that, although they infrequently occur (only 6\% of all HOs are to 2G/3G), HOs to 3G (2G) increase the HOF rate by 166\% (915\%) compared to HOs between the newer RATs (intra 4G/5G-NSA).

\rev{\noindent\textbf{Organization.} The remainder of this paper is organized as follows. \cref{sec:mobility_management} covers relevant handover concepts, while \cref{sec:dataset_methodology} details the datasets and metrics used in our study. \cref{sec:first_look_at_network} introduces the three main axes of analysis that we use in the paper, and \cref{sec:mobility_analysis} shares insights on HOs based on these axes; \cref{sec:areas_of_attention} examines the impact and causes of HOFs. Additionally, \cref{sec:related_work} reviews related literature, providing context for our findings. Lastly, \cref{sec:limitations} and \cref{sec:conclusion} discuss the study's limitations and avenues for future work, and conclude with a summary of the main findings.}

\section{Background}
\label{sec:mobility_management}

\begin{figure}[t]
    \centering
    \includegraphics[scale=0.31]{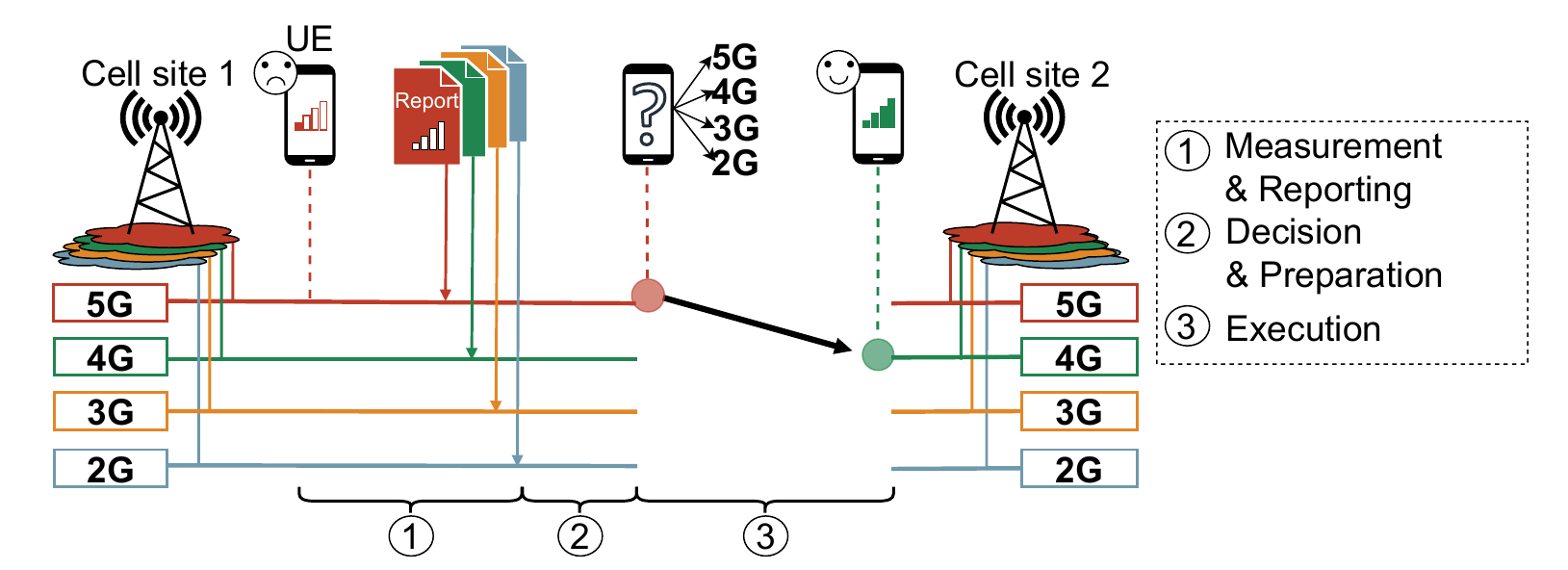}
    \vspace{-6mm}
     \caption{High-level description of HO procedure~\cite{3gpp_23_401}.}
     \label{fig:ho_steps}
     % \Description{High-level description of HO procedure.}
\end{figure}

MNOs install \textit{cell sites} in various locations to ensure widespread service availability. These sites are typically equipped with multiple antennas associated with \textit{radio sectors} (or simply, \textit{sectors}) that support different RATs and serve the UEs residing in a bounded geographical area \cite{stuber2001principles}. This design principle mandates that, as UEs move, they must transition, or \textit{handover}, across different sectors. To offer a seamless network connectivity experience for mobile users, it is essential for HOs to execute timely and reliably.\footnote{The HO procedure is different from \textit{cell (re-)selection}, which happens when users do not maintain an active data connection but still need to change sectors to ensure the reception of signaling messages \cite{survey_ho}.}

Configuring the cell sites and their sectors to optimally handle HOs is crucial: frequent HOs may lead to excessive signaling and unneeded service interruptions, whereas scarce HOs might result in poor signal reception or temporary lack of connectivity. Moreover, with the advent of 5G and its two deployment strategies \cite{SA_vs_NSA}, NSA and SA, HO management has become more intricate \cite{5G_wild_2021} due to the coexistence and integration of multiple generations of mobile technologies;\footnote{From a mobility management perspective, 5G-NSA and 4G are identical, since the former relies on the 4G Evolved Packet Core (EPC) functions.} at the same time, a crucial goal of 5G-SA is to provide support for Ultra Reliable Low Latency Communications (URLLC), which mandates the provision of reliable and fast HOs.

\noindent\textbf{Handover Mechanism.} Every UE relies on its \textit{primary sector} (i.e., the sector it is connected to), serving as the pivotal link for control-plane signaling and HO management. Figure \ref{fig:ho_steps} depicts the HO process from a source (i.e., primary) to a target radio sector~\cite{3gpp_36_413}. When a UE attaches to a new sector, it receives a set of mobility management configurations, including parameters for the triggering of HO events (e.g., hysteresis, offsets, etc.). Based on these configurations, the UE performs signal strength/quality measurements --~e.g., Reference Signal Received Quality (RSRQ)~-- of the source and neighboring sectors, and sends a Measurement Report~(MR) to the source periodically, or if any of the mobility management criteria is met. For instance, in 4G and 5G NR, a HO triggering event typically occurs when either the serving sector's signal falls below a threshold (A2 event) or when the signal of a neighboring sector becomes offset better than the serving sector (A3 event)~\cite{3gpp_36_331, 3gpp_38_331}. Based on the MR, the source identifies the best target sector and initiates the HO. After the target sector accepts the request, the source transmits a HO command to the UE. For example, in 4G/5G NR, the source sends a Radio Resource Control (RRC) Connection Reconfiguration message to the UE to begin its cell synchronization with the target sector and the Random Access Channel (RACH) procedure. After the UE reports successful access to the target sector, the source releases its resources. More details are available in~\cite{3gpp_23_401, mollel2021survey}.

\section{Datasets \& Methodology}
\label{sec:dataset_methodology}

In this section, we present our measurement infrastructure in a large MNO in the \revtwo{studied country.} We explain the three datasets we built for this study and introduce the \revtwo{official census dataset} we used to complement our analysis. Finally, we detail the employed performance and mobility metrics.

\subsection{Network Data Collection}\label{subsec:data-feeds}

\begin{figure}[t]
    \centering
    \includegraphics[scale=0.3]{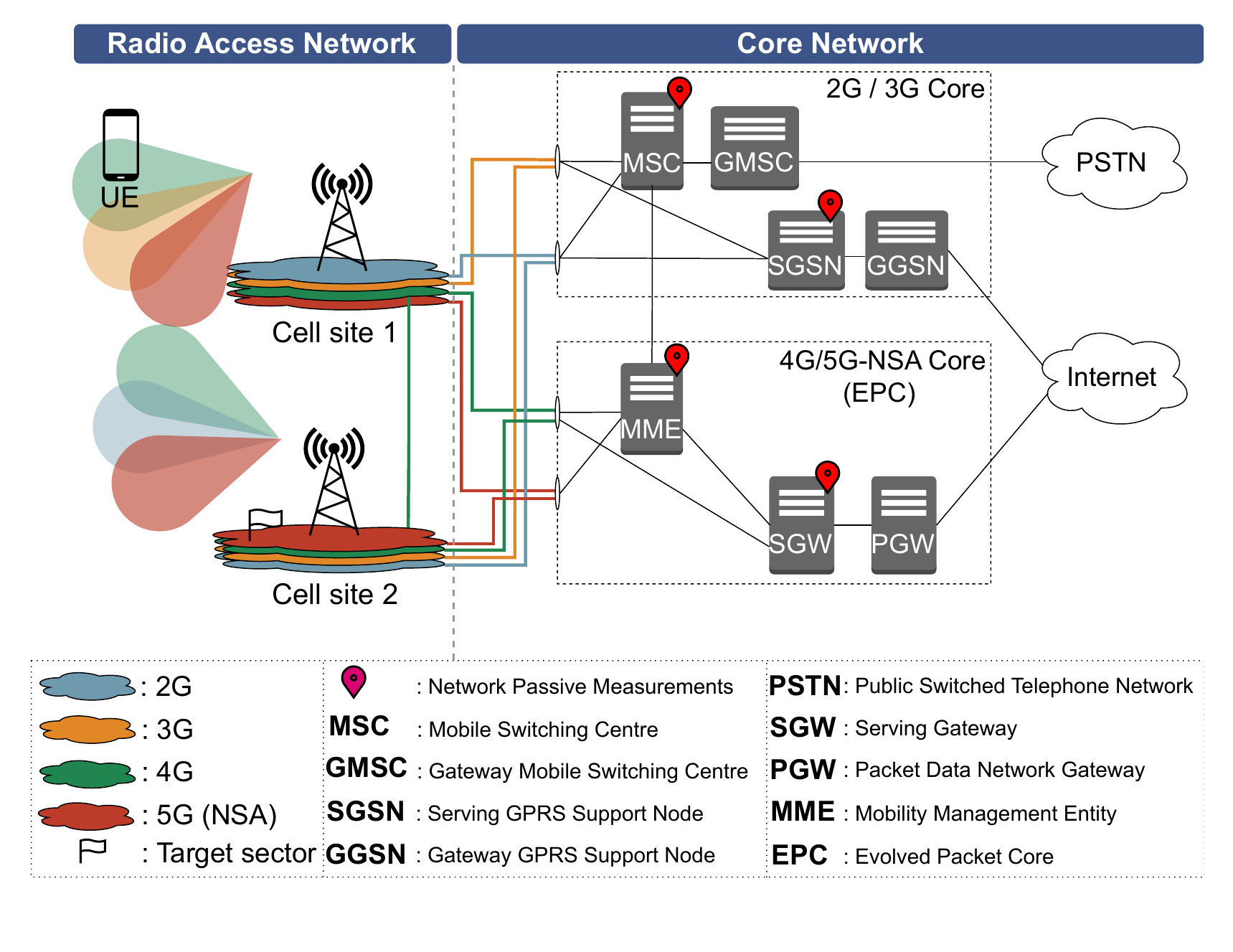}
    \vspace{-10mm}
    \caption{Network architecture \& measurement points.}
    \label{fig:HO-archi}
    % \Description{Network architecture and measurement points.}
    % \vspace{-5mm}
\end{figure}

\textbf{Measurement Infrastructure.} We collect passive measurements using commercial tools integrated into the MNO's infrastructure, see Figure \ref{fig:HO-archi}. In a nutshell, the cellular network architecture can be divided into three primary components: ($i$)~the devices accessing the network, ($ii$)~the Radio Access Network (RAN), responsible for managing wireless communication, and ($iii$)~the Core Network (CN), which provides the required services and functions for the network operation (e.g., user authentication and mobility management). This is consistent for all the different radio technology generations that coexist in the network. Our monitoring locations, which we depict with red pins in Figure~\ref{fig:HO-archi}, focus on key components of the core network, including the Mobile Management Entity (MME tracks and manages the mobility of devices in 4G and 5G-NSA), the Mobile Switching Center (MSC is responsible for communication switching functions), the Serving GPRS Support Node (SGSN manages data routing for 2G/3G), the Serving Gateway (SGW routes packages between RAN and the CN), and the cell sites in the RAN.

\revthree{Data is collected in a private cloud environment for a given retention period, and is already anonymized before we process it. Particularly,} we organize the collected data into three datasets, providing various information at the radio sector and UE-level.

\noindent\textbf{Mobility Management Signaling Dataset.} The captured data spans from 29-Jan-2024 to 25-Feb-2024 for the entire \revtwo{country} (see Table~\ref{tab:statistics}). We analyzed the activity of users in the control plane for all RATs supported by the MNO. For each RAT, the dataset includes the (control plane) signaling messages related to events such as service requests, HOs, attach/detach, paging, and Tracking Area Update (TAU). We direct our attention to HOs, for which we capture six main variables that enable an \mbox{in-depth} analysis: ($i$)~\textit{timestamp}, with millisecond granularity, ($ii$)~\textit{HO result} (i.e., success/failure), ($iii$)~\textit{HO duration} (msec granularity), ($iv$)~\textit{cause codes for HO failures based on 3GPP}~\cite{3gpp_36_413, 3gpp_29_274}, which are enriched with \textit{sub-cause descriptions} specified by the antenna vendors, ($v$) \textit{anonymized user ID}, based on the International Mobile Subscriber Identity (IMSI) and the International Mobile Equipment Identity (IMEI),\footnote{The first 8 digits of the IMEI represent the Type Allocation Code (TAC), which we use later to classify devices.} and ($vi$) source and target radio sectors along with their RATs. As mentioned before, due to the early stages of 5G-SA deployment in the studied MNO, we base our analysis on 5G-NSA\revtwo{, which relies on the 4G EPC for mobility management.}

\noindent\textbf{Radio Network Topology.} We utilize this dataset to integrate in our analysis the upgrades in the MNO's network deployment footprint (e.g., newly deployed sites). We capture this dataset daily during the period of analysis; it contains information on each deployed radio sector, such as geographic location (longitude and latitude), the postcode of the area, and the supported technologies (i.e., 2G, 3G, 4G, 5G).

\noindent\textbf{Devices Catalog.} We leverage a daily commercial database, provided by the Global System for Mobile Communications (GSM) Association (GSMA) to examine correlations of device-specific characteristics with HOs. This catalog associates the TAC of each device with attributes such as the supported radio bands and RATs, the manufacturer, and the device type. We apply a heuristic to classify the devices into three types: smartphones, M2M/IoT devices, and low-tier feature phones~\cite{lutu2020things}. For this, we rely on the observation that the Access Point Name (APN) configured for the UEs may contain keywords associated with IoT verticals (e.g., ``m2m'', ``smart-meter''), and combine the information from the APN with the device characteristics of our daily commercial GSMA database.

\subsection{\revtwo{Census Dataset}}\label{subsec:census-dataset}

We leverage open datasets published by \revtwo{the official census office in the studied European country} to enrich our mobility study with the geodemographic characteristics \revtwo{of different areas.} \revtwo{Specifically, we take as a reference the 300+ districts defined by the census office, and collect the population density and the postcodes within each of them. Then, based on census information we classify postcode areas into two main categories (\textit{urban} and \textit{rural}), which correlate with population density (more than 10k and less than 10k residents) and also serve as a proxy for areas with denser and sparser RAN deployments, respectively.}

\subsection{Performance \& Mobility Metrics} \label{sec:evaluation-metrics}

\textbf{Performance Metrics.} In line with prior works \cite{survey_ho, ho_problems}, we focus on the following metrics.

\noindent$\bullet$ \textit{HO count} represents the number of HOs over a time interval, which we usually set to either 30 minutes, 60 minutes, or one day.
We use this metric to show how users' mobility fluctuates over time and space, and how it differs per RAT, device type, and manufacturer. 

\noindent$\bullet$ \textit{HO duration} represents the time interval (in msec) to complete the HO, see \cite{wheels_2023, ho_times, ho_times_2, ho_times_3}. Minimizing this interval is crucial for seamless connectivity and improves the users' Quality of Experience (QoE).

\noindent$\bullet$ \textit{HOF rate} refers to the number of HO failures (HOFs) divided by the total number of triggered HOs.\footnote{We primarily focus on HO failures rather than explicitly detailing HO successes; however, successes and failures are complementary to each other.} HOFs dramatically affect the users' experience and typically happen due to poor signal strength, configuration and synchronization errors, or capacity issues in the network. In \cref{sec:areas_of_attention}, we uncover the reasons why these failures occur and emphasize that a comprehensive understanding of HOFs can only be achieved by incorporating the perspective of MNOs.

\noindent\textbf{Mobility Metrics.} To showcase the mobility characteristics of users, we focus on two metrics from the MNO's perspective, as follows. 

\noindent$\bullet$ \textit{Number of sectors} quantifies the number of distinct radio sectors that a user successfully communicates with, per day. We highlight that this metric does not necessarily translate to the distance traveled by users in a given area, as the density of radio sectors in the area also plays a role. For instance, urban areas typically have denser deployments and, as a result, users connect to a larger number of sectors even if they travel the same distance as in less populated areas (e.g., rural) with sparser deployments.

\noindent$\bullet$ \textit{Radius of gyration} complements the previous metric by capturing the traveled distance for the UEs~\cite{González2008_gyration}. It is defined as the root mean squared distance between each visited sector (weighted by the time spent there) and the center of mass. \rev{The radius of gyration is defined as follows: 
\begin{equation*}
g=  \sqrt{\frac{1}{N} \sum_{j=1}^{N} (t_j\textbf{l}_j - \textbf{l}_{cm})^2},  
\end{equation*}
where $\textbf{l}_j$ represents the location of the $\textit{j}^{th}$ visited cell site, $t_j$ represents the time spent in the $\textit{j}^{th}$ visited cell site and $\textbf{l}_{cm}$ represents the location of the user's center of mass, calculated as $\textbf{l}_{cm} =  \frac{1}{N} \sum_{j=1}^{N} \{ t_j\textbf{l}_j$\}, where $N$ is the total number of cell sites visited by the user.} A high radius of gyration indicates that the user travels far and wide (i.e., their moves span a large geographical area). Conversely, a lower radius of gyration points to more localized movements, relatively close to a central location.

\section{Exploring Data Heterogeneity} 
\label{sec:first_look_at_network} 

Our datasets capture the heterogeneity and complexity of HOs across the entire MNO's deployment in the \revtwo{studied country,} which includes diverse deployment densities and RATs, as well as a broad spectrum of UEs (e.g., smartphones, M2M/IoT, etc). In this section, we explore the heterogeneity of these datasets along three particularly interesting axes, from the network's perspective: $(i)$~heterogeneity of RATs, $(ii)$~heterogeneity of UEs, and $(iii)$~geodemographic complexity.

\subsection{RAT Deployment \& Usage}
\label{subsec:heterogeneity-RATs}

\begin{figure}[t]
    \centering
    \begin{subfigure}[H]{0.49\columnwidth}
         \centering
         \includegraphics[width=\columnwidth]{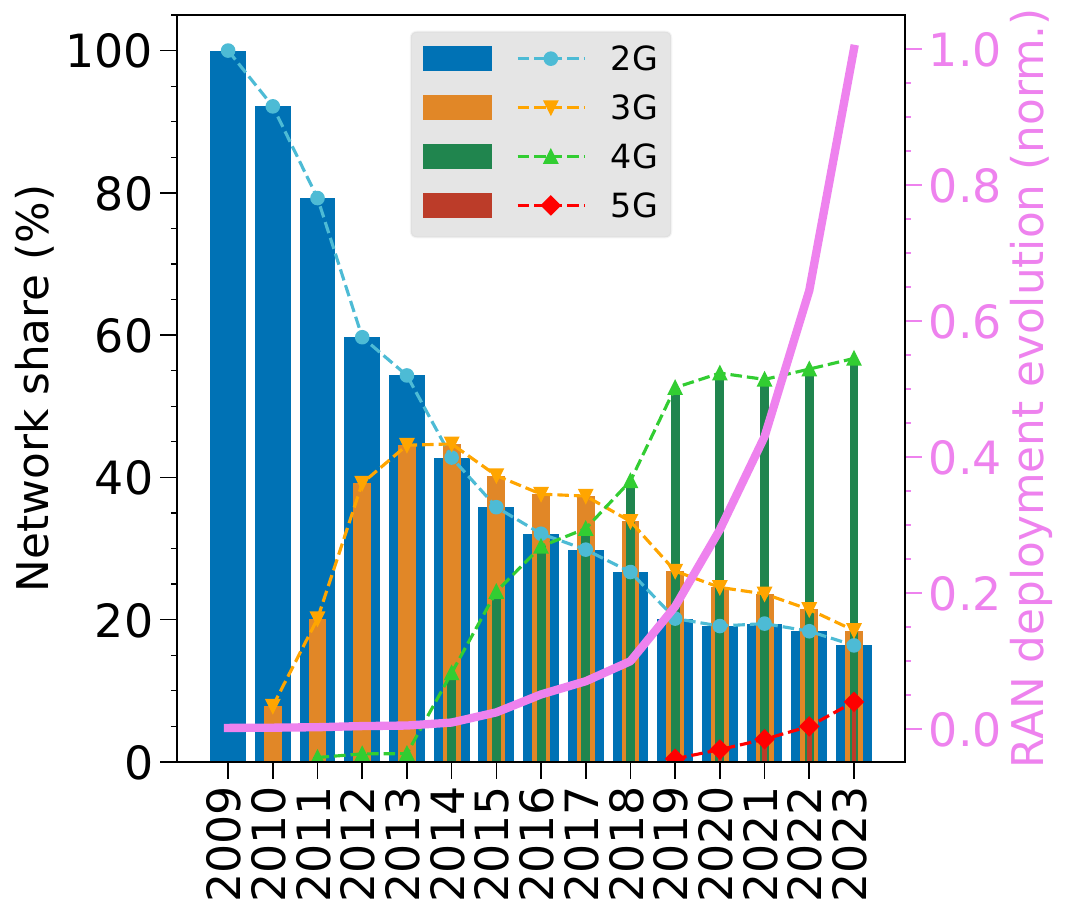}
         \caption{}
         \label{fig:num_cells}
     \end{subfigure}
     \begin{subfigure}[H]{0.49\columnwidth}
         \centering
         \includegraphics[width=\columnwidth]{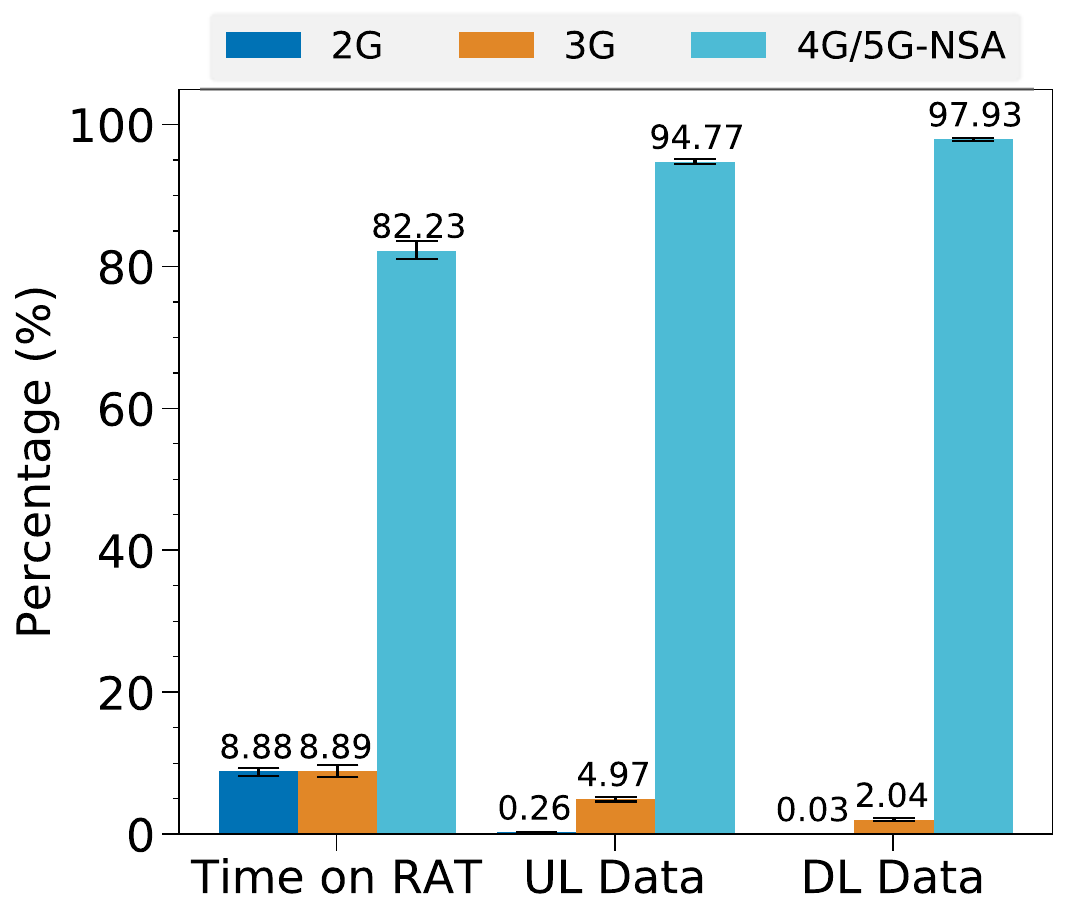}
         \caption{}
         \label{fig:usage-stat-RATs}
     \end{subfigure}
     % \vspace{-3.5mm}
     \caption{(a) Evolution of network deployment in a commercial MNO. The left y-axis 
     corresponds to all bars and lines, except the pink line (right y-axis). (b) Average daily RAT use. Error bars show the min/max daily values over 4 weeks.}
     % \Description{(a) Evolution of network deployment in a commercial MNO. The left y-axis 
     % corresponds to all bars and lines, except the pink line (right y-axis). (b) Average daily RAT use. Error bars show the min/max daily values over 4 weeks.}
     % \vspace{-4mm}
\end{figure}

Figure~\ref{fig:num_cells} shows the deployment evolution in the network from 2009 to 2023. The number of sectors (solid pink line) has increased at an exponential pace in the last 15 years, with an average growth of 59\% during the last 5 years (2018-2023). Throughout these 15 years, different RATs have been coexisting with a varying mix. The latest major network upgrade occurred in 2019 with the deployment of 5G-NR, which accounted for 8.4\% of the sectors by the end of 2023. At the same time, we observe the gradual decommissioning of 2G and 3G cells ($\approx$18\% each in 2023), while 4G is still the dominant RAT ($\approx$55\%) in terms of infrastructure. This heterogeneity does not come as a surprise, since decommissioning legacy RATs is a challenging process that needs to account for various techno-economic factors, such as the turnover rate of customers or the radio coverage \cite{ericcson_sunset2G3G}. Nonetheless, it compounds network management and affects both the number and the success of HOs as we present in \cref{sec:areas_of_attention}, and as prior studies have also identified~\cite{li2021nationwide, yuan2022understanding}.

To further understand the use of the RATs, we compute the overall time that UEs spend on each of them by using the timestamps of mobility events in the dataset. With the current 5G-NSA deployment, we do not distinguish from the events captured in the core network (i.e., MME) when devices are served by a 4G or a 5G-NR sector (see \cref{sec:mobility_management}); thus, we use the term ``4G/5G-NSA''. In Figure~\ref{fig:usage-stat-RATs}, we notice that UEs rely mostly on 4G/5G-NSA ($\approx$82\% of the time on average), while 2G and 3G serve users during a non-negligible 8.9\% of the time each. In terms of aggregated data volumes, the share for 4G/5G-NSA rises up to 94.77\% and 97.93\%, respectively, for UL and DL traffic, leaving marginal values for 2G and 3G. Yet, these legacy RATs still serve a noteworthy number of UEs that support only these technologies (see \cref{subsec:heterogeneity-devices}).

The heterogeneity of the network appears also in terms of the antenna vendor. Four principal vendors (V1, V2, V3, V4) employ antennas (and thus, RATs) for this network, with their deployment distributed asymmetrically across 
different regions. All vendors support 4G/5G-NSA and 3G RATs, and accommodate nearly the full spectrum of devices. Details are provided in Appendix \ref{sec:appendix-regression}.

\noindent\textbf{\textit{Key takeaways}}: \textit{The cellular network we measure includes all RATs (2G-5G), where 2G and 3G radio sectors account for 36\% of the total deployment. These RATs (2G \& 3G) connect users on average for 18\% of their up-time, while UEs generate only 5.23\% (2.07\%) of the UL (DL) data over them.}

\begin{figure}[t]
    \centering
    \begin{subfigure}[H]{0.48\textwidth}
         \centering
         \includegraphics[width=\columnwidth]{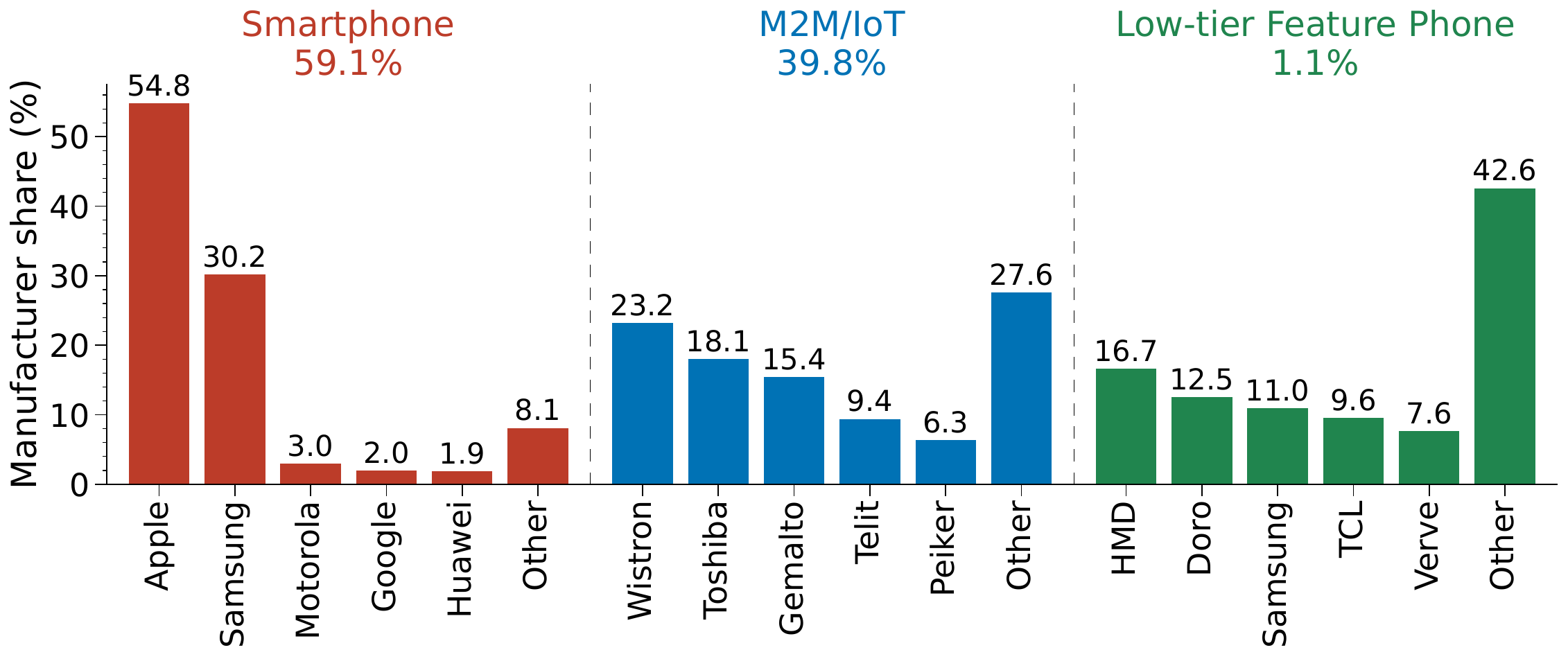}
         \caption{Manufacturers} 
         \label{subfig:manufacturers}
     \end{subfigure}
    \begin{subfigure}[H]{0.48\textwidth}
         \centering
         \includegraphics[width=\columnwidth]{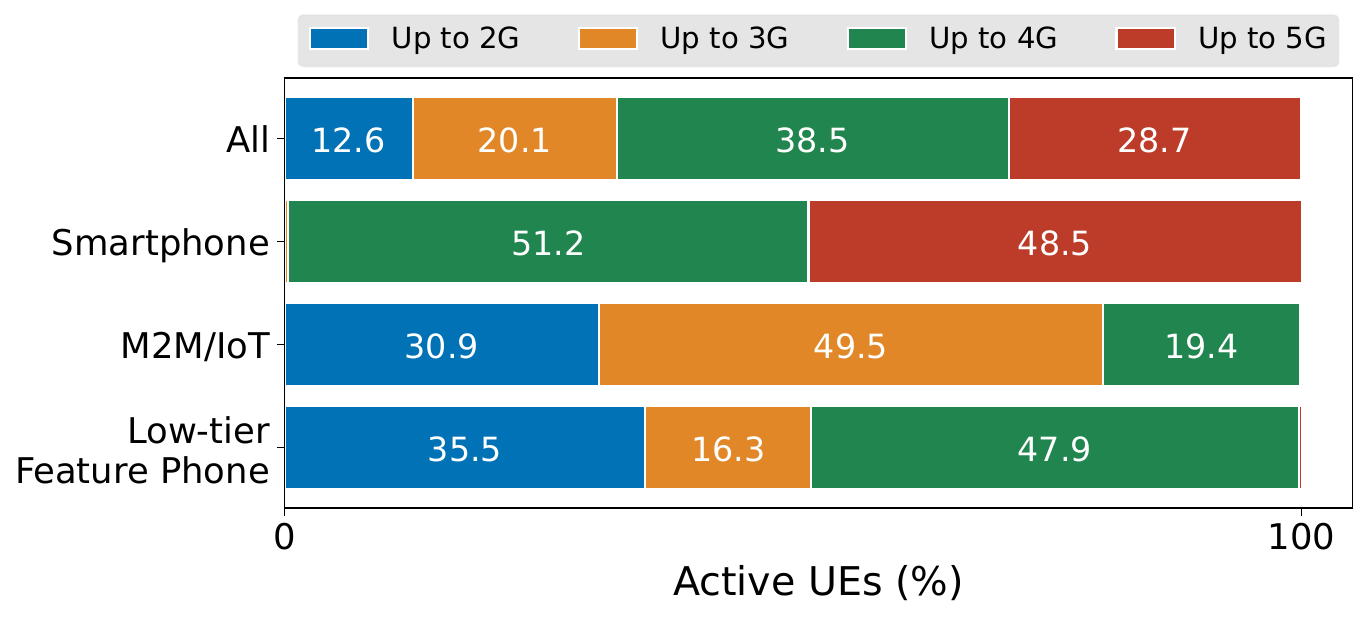}
         \caption{RAT support}
         \label{subfig:uptoG}
     \end{subfigure}
     % \vspace{-3mm}
     \caption{Percentages of (a) manufacturers and device types, and (b) the RATs they support (excluding <0.5\%).
     % \Description{Percentages of (a) manufacturers and device types, and (b) the RATs they support (excluding <0.5\%).}
     }
     \label{fig:het_rat_man}
     % \vspace{-4mm}
\end{figure}

\subsection{User Equipment}
\label{subsec:heterogeneity-devices}

The number of devices accessing the network over the 4-week period is $\approx$40M. We classify these devices into three types based on their capabilities, namely, \textbf{smartphones}, \textbf{M2M/IoT devices}, and \textbf{low-tier feature phones}, accounting for 59.1\%, 39.8\%, and 1.1\% of UEs, respectively. Figure~\ref{subfig:manufacturers} shows the top-5 manufacturers in the three types of devices. In the larger \revtwo{category} -- smartphones -- we observe that most devices are manufactured by Apple (54.8\%) or Samsung (30.2\%). For M2M/IoT UEs, we find a diversified set of manufacturers; namely, over 27\% of these UEs are from manufacturers outside the top-5. 

We infer the connectivity capabilities of mobile devices from the GSMA device catalog (see \cref{sec:dataset_methodology}). We find that 12.6\% of all UEs support only 2G and 20.1\% up to 3G (see Figure \ref{subfig:uptoG}), which partially explains the slow pace of decommissioning legacy RATs. These legacy devices are mostly M2M/IoT devices or feature phones, where $>\!80\%$ and $>\!50\%$, respectively, support at most 3G. The overall number of devices that support 4G or 5G adds up to 67.2\%. The majority of these devices are smartphones: 51.4\% of smartphones support up to 4G, and 48.5\% are 5G-capable.

\noindent\textbf{\textit{Key takeaways}}: \textit{Over 32\% of all devices support only up to 3G -- predominantly M2M/IoT UEs and feature phones -- and 51.5\% of smartphones do not support 5G yet (the majority relies on 4G). These factors contribute to the presence of a mixture of old and new RATs in current deployments, stressing the challenges associated with decommissioning the older ones.}

\subsection{Geodemographic segmentation}

\begin{figure}[!t]
    \centering
    \includegraphics[width=\columnwidth]{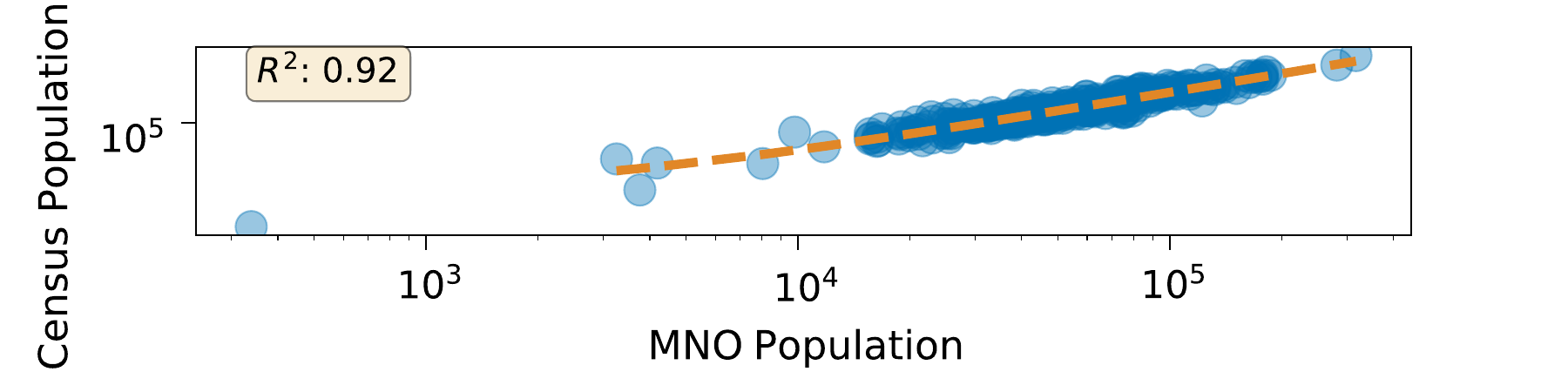}
    % \vspace{-5mm}
     \caption{Comparison between the inferred population from the MNO data and the actual population from the census data (\revtwo{district}-level).
     }
    % \Description{Comparison between the inferred population from the MNO data and the actual population from the census data (district-level).}
     % \vspace{-5mm}
     \label{fig:repr}
\end{figure}

\textbf{Population Sampling.} 
This section demonstrates that the dataset we collect through the commercial MNO is representative of the \revtwo{country's overall population.} Figure~\ref{fig:repr} shows the population according to census (y-axis) and the population we inferred from the MNO (x-axis), where each data point refers to the \revtwo{districts in the country} (see \cref{sec:dataset_methodology}). We derive the end-user's home location at postcode granularity from their connectivity patterns during nighttime~\cite{home_detection_2010}. To achieve this, we consider the main cell site the user connects to between 00:00 and 08:00 (i.e., night hours) for at least 14 days (not necessarily consecutive) during February 2024. We then aggregate their mapped home postcode at the \revtwo{district} level. These results show a clear linear relationship ($R^2 = 0.92$) between the census data and the MNO user base, which reinforces that our dataset \revtwo{accurately} captures the \revtwo{country's} population distribution~\cite{phithakkitnukoon2012socio}. This renders our dataset especially interesting for analyzing mobility in the entire country, including regions with diverse population dynamics.

\begin{figure}[t]
    \centering
    \includegraphics[width=0.9\columnwidth]{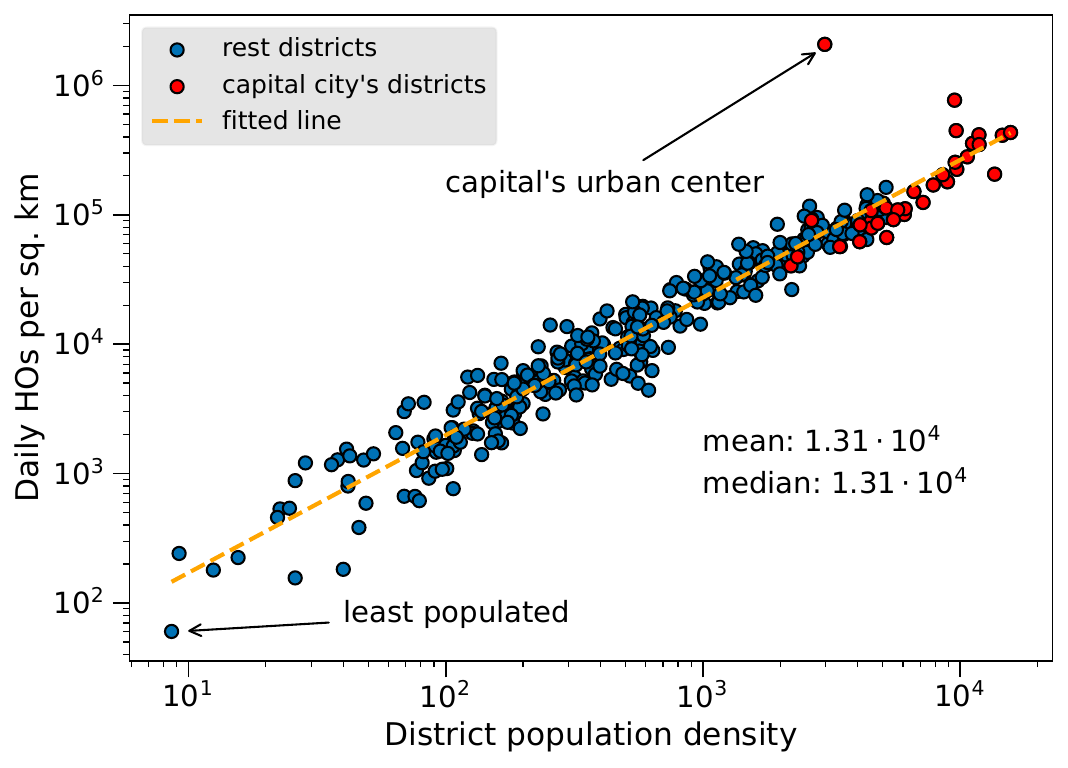}
    % \vspace{-4mm}
     \caption{Daily HOs per sq. km in the
     \revtwo{country (district level).}}
     \label{fig:num_hos_lads}
     % \Description{Daily HOs per sq. km in the country (district level).}
     % \vspace{-5mm}
\end{figure}

\noindent\textbf{Mobility \& Geodemographics.} 
We investigate the distribution of mobility events by examining the number of HOs across \rev{districts.} Figure~\ref{fig:num_hos_lads} shows \revtwo{the number of daily HOs per sq. km in each district, together with the population density there (residents per sq. km).} This analysis facilitates the characterization of mobility patterns across distinct geodemographic segments (e.g., densely populated urban areas or less populated rural areas). Overall, our findings indicate a strong positive correlation (Pearson correlation of 0.97), between the number of HOs \revtwo{per day} and the residential population \revtwo{density} in the corresponding \revtwo{district.}

As anticipated, dense urban areas exhibit a high number of HOs per square km. For instance, in the \revtwo{district that covers the urban center of the capital,} we observe approximately 2.1M HOs per square km each day. In this city the studied network's infrastructure itself comprises more than 500 radio sectors per square km. Conversely, in less populated areas the intensity of HOs is significantly lower (60 HOs per sq. km \revtwo{in the least densely populated district}). This value is more than 200$\times$ lower compared to the \revtwo{district-level} mean \revtwo{in the country} (13.1k HOs per sq. km daily), reflecting the stark contrast in mobile network activity between highly urbanized and more remote areas.

\noindent\textit{\textbf{Key takeaways}: We infer the home locations of approximately 40M UEs across the \revtwo{studied country} to ensure that our data accurately represents the entire population ($R^2 = 0.92$ with census data). By analyzing HOs per square km at the \revtwo{district level,} we observe significant disparities -- from 2.1M daily HOs per sq. km in the \revtwo{in the center of the capital city} to 60 HOs \revtwo{per sq. km} in remote areas -- highlighting the complexity of managing HOs across different regions.}

\section{Characteristics of Handovers}
\label{sec:mobility_analysis}

Analyzing mobility patterns is crucial for various purposes, including urban planning, social policy design, and optimizing network infrastructure~\cite{zhang2018mobility, li2020beyond}. In this section, we take as reference the three axes of heterogeneity from \S\ref{sec:first_look_at_network}, and characterize geo-temporal cellular mobility patterns through HOs. We examine the horizontal and vertical HOs across UE types and \revtwo{districts,} and investigate how mobility and UE manufacturers relate to HO performance.

\subsection{Geo-temporal Analysis}\label{sec:mob_traffic_pattern}

First, we analyze HO patterns as a function of geodemographic factors, focusing on the difference between urban and rural \revtwo{areas, classified at the postcode level (\S\ref{subsec:census-dataset})}.\footnote{\revtwo{We drop from this analysis 3.1\% of the postcodes due to the lack of reliable census information in these areas.}} \revtwo{This broader urban/rural classification enables us to robustly capture variations in HO dynamics across areas with different demographic characteristics.}

\begin{figure}[t]
    \centering
    \includegraphics[width=\columnwidth]{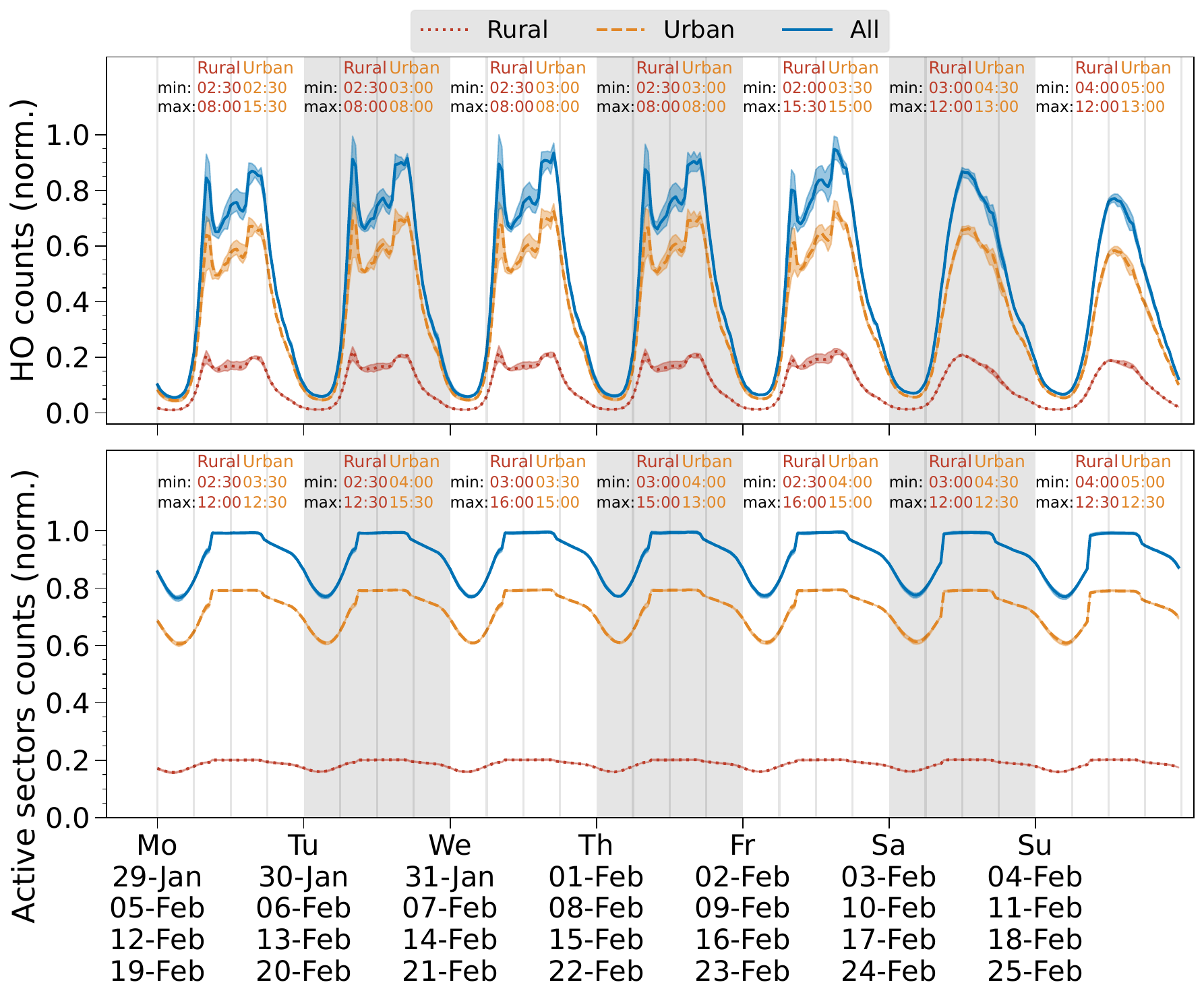}
    % \vspace{-5mm}
     \caption{Temporal evolution of HOs (top) and active sectors (bottom) in urban and rural areas. Curves show the average HO volume in 30-minute intervals over the four weeks; shadows show the min/max values. All values are normalized by the max HO and sector count seen over the period.} 
     % \Description{Temporal evolution of HOs (top) and active sectors (bottom) in urban and rural areas. Curves show the average HO volume in 30-minute intervals over the four weeks; shadows show the min/max values. All values are normalized by the max HO and sector count seen over the period.}
     \label{fig:ho_and_sectors_per_day}
     % \vspace{-1mm}
\end{figure}

\noindent\textbf{HO Patterns.} Figure~\ref{fig:ho_and_sectors_per_day} (top) shows the weekly temporal evolution (with \textit{30-min granularity}) of HO counts over the 4-week period (shadows show the min/max values). To adhere to privacy and security guidelines of the MNO, we normalize the HO counts by the max value (over 30-min intervals) of the studied period. The total number of HOs in urban areas represents on average 78\% of all HOs, which is consistent with findings from other studies \cite{pp_ho_2023}. Namely, we discover that 80\% of the sectors are installed in urban areas, which cover only 49.6\% of the total \revtwo{territory of the country.}

From the daily HO patterns, we observe that weekdays (Mo-Fr) experience higher number of HOs compared to weekends (Sa-Su). Concretely, we find a 33\% reduction on average in the peak of HOs during Sundays compared to \rev{Fridays}. Moreover, we identify the peak HO times during weekdays at 08:00--08:30 and 15:00--15:30 for both rural and urban areas. Also, weekday HO patterns exhibit notable fluctuations, with a sharp $\times$3 increase in the HOs observed from 06:00 to 08:00; this is in contrast to weekends, which have a single peak of mobility between 12:00 and 13:00.
During weekdays, after the second peak at 15:00--15:30, the number of HOs gradually decreases (on average 11\% per 30 minutes), leading to the minimum at 02:00--03:30 (or 03:00--05:00 over the weekends).

Likewise, Figure~\ref{fig:ho_and_sectors_per_day} (bottom) shows the number of active sectors --~handling at least one HO~-- over 30-min intervals. As underlined in the sequel, MNOs apply dynamic energy-saving policies to switch off sectors when they are not needed to satisfy capacity demand. Comparing Figures~\ref{fig:ho_and_sectors_per_day} (top) and~\ref{fig:ho_and_sectors_per_day} (bottom), we see that the portion of active sectors highly correlates with the HO counts (Pearson correlation of 0.9). Weekdays and weekends present no significant differences in terms of active sectors. More precisely, after 08:00 (first peak hour) $\approx$99\% of sectors remain active until 17:00, when a decrease of $\approx$1\% per 30-min is observed, until midnight. As mentioned earlier, we conjecture that this decrease correlates not only with the reduced mobility of the UEs (notice the HO drop at the same hours), but also with the reduced capacity demand (i.e., less user activity) in densely deployed areas, which triggers energy-saving mechanisms to switch off sectors that act as capacity boosters~\cite{tan2022energy, de2023modeling}.

\noindent\textbf{\textit{Key takeaways}}: \textit{Handover patterns vary significantly across: $(i)$ urban and rural areas (urban sectors account for 78\% of HOs, while covering only 49.6\% of the territory), and $(ii)$ during weekdays and weekends (33\% of difference during peak hours). }

\subsection{Horizontal vs Vertical HOs}\label{ho-types}

To understand how devices interact with the different RATs in the network (see \S\ref{sec:first_look_at_network}), we take as a reference the behavior of devices connected to 4G/5G-NSA (i.e., 4G and 5G-enabled devices). We differentiate three main \textit{HO types}, namely, intra 4G/5G-NSA (horizontal), 4G/5G-NSA~$\rightarrow$~3G (vertical), and 4G/5G-NSA~$\rightarrow$~2G (vertical). Our intent is to characterize how frequently these devices still rely on older RATs, and in which circumstances.

\noindent\textbf{HO Frequencies.} Table~\ref{tab:horizontal_vs_vertical_hos} depicts the percentage of the different HO types we registered across UE types. The vast majority of HOs are intra 4G/5G-NSA HOs (94.14\%), while vertical HOs --~from 4G/5G-NSA to 3G or 2G~-- correspond to 5.86\% and 0.001\%, respectively.

Furthermore, smartphones primarily initiate intra-4G/5G-NSA HOs, contributing to 88.28\% of the total, with a fallback to 3G occurring in 5.84\% of the cases. M2M/IoT devices engage mostly in intra 4G/5G-NSA HOs, with a minority transitioning to 3G, a pattern echoed by feature phones on a smaller scale (i.e., 0.13\% intra 4G/5G-NSA HOs). This is particularly important given that $\approx$80\% of M2M/IoT devices support only up to 3G (see Figure~\ref{subfig:uptoG}). It is an artifact of the IoT vertical applications employing massive M2M deployments (e.g., smart meter applications), which often require only stationary devices with limited connectivity demands~\cite{lutu2020things}.

\begin{table}[t]
    \centering
    \caption{Statistics per handover and device type.}
    \vspace{-2mm}
    \resizebox{1.0\columnwidth}{!}{
    \begin{tabular}    {c@{\hspace{0.1mm}}c@{\hspace{2mm}}c@{\hspace{2mm}}c@{\hspace{2mm}} c}
    \toprule
    & \textbf{Horizontal} & \multicolumn{2}{c}{\textbf{Vertical}} & \multirow{2}{*}{\shortstack{ \textbf{All} \\ \textbf{HOs (\%)}}} \\
    \cmidrule{2-4}
    & \makecell{\textbf{\small{Intra 4G/}} \\ \textbf{\small{5G-NSA (\%)}}} & 
    \makecell{\textbf{\small{4G/5G-NSA}} \\ \textbf{\small{to 3G (\%)}}} & \makecell{\textbf{\small{4G/5G-NSA}} \\ \textbf{\small{to 2G (\%)}}} & \\
    \midrule
    \textbf{\small{Smartphones}} & $88.28\pm 0.77$ & $5.84\pm 0.77$ & $<0.001$ & $94.12\pm 0.77$ \\
    \textbf{\small{M2M/IoT}} & $5.73\pm 0.52$ & $0.02\pm 0.01$  & $<0.001$ & $5.75\pm 0.53$\\
    \textbf{\small{Feature phones}} & $0.13\pm 0.05$ & $<0.001$  & $<0.001$ & $0.13\pm 0.05$\\
    \textbf{\small{All devices}} & $94.14\pm 1.29$  & $5.86\pm 0.78$  & $<0.001$ & - \\
    \bottomrule\bottomrule
    \end{tabular}
    }
\label{tab:horizontal_vs_vertical_hos}
\end{table}

\begin{figure}[!t]
    \centering
    \includegraphics[width=\columnwidth]{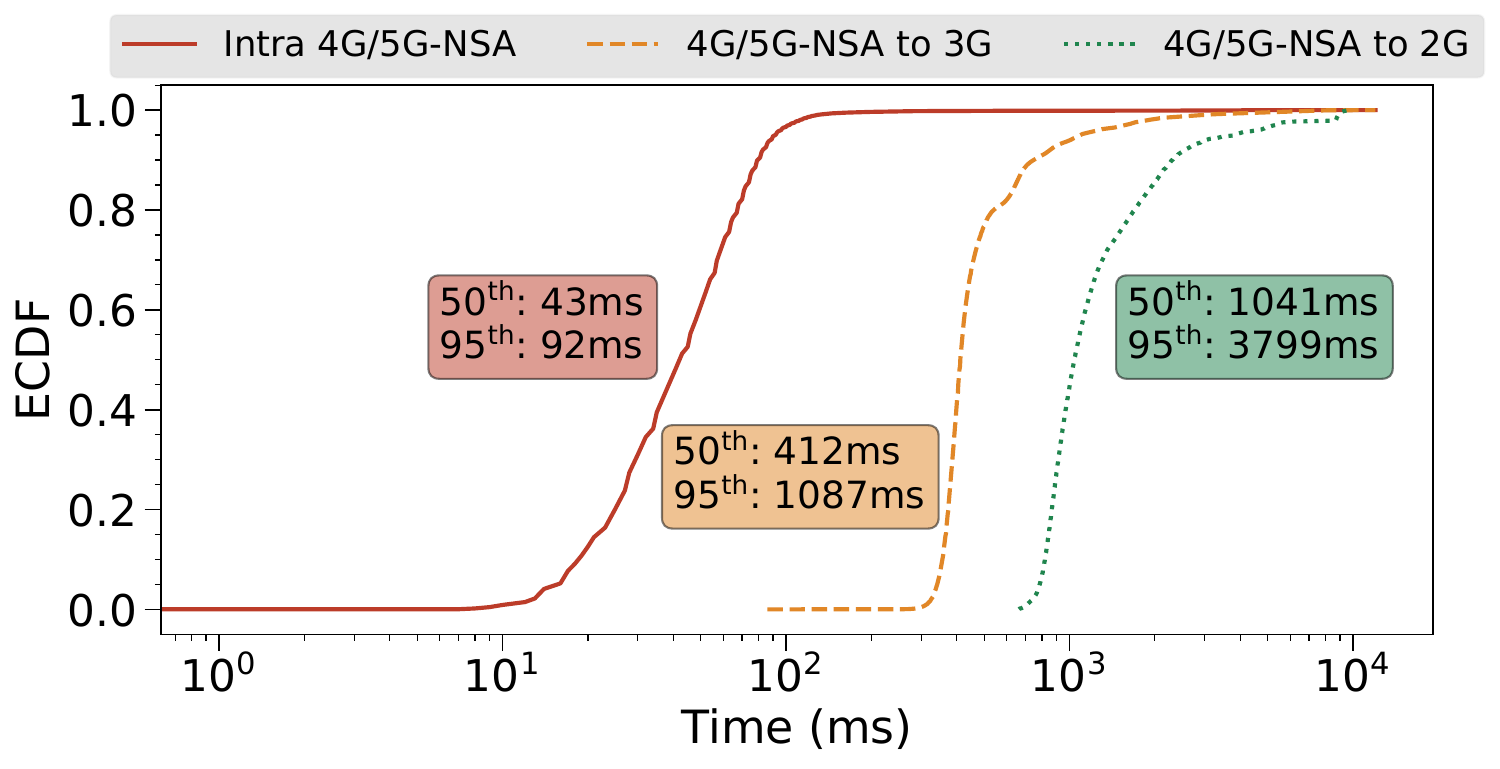}
     \caption{HO duration (horizontal vs vertical).}
     \label{fig:ecdf_ho_time}
     % \Description{HO duration (horizontal vs vertical).}
     % \vspace{-4mm}
\end{figure}

\begin{figure*}[t]
    \centering
    \begin{subfigure}[H]{0.25\textwidth}
         \centering
         \includegraphics[width=\textwidth]{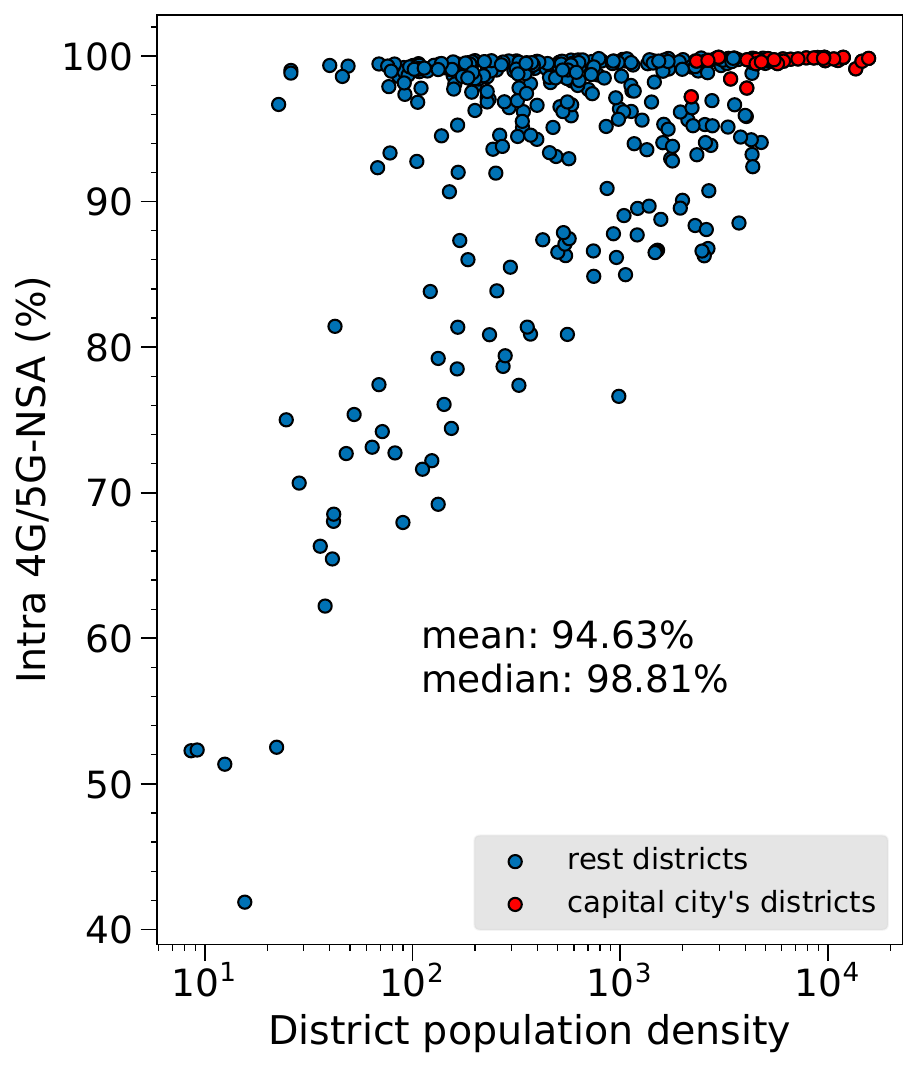}
         % \vspace{-2mm}
         \caption{}
         \label{subfig:heatmaps_rat_to_rat_intra}
     \end{subfigure}
     \hfil
     \begin{subfigure}[H]{0.25\textwidth}
         \centering
         \includegraphics[width=\textwidth]{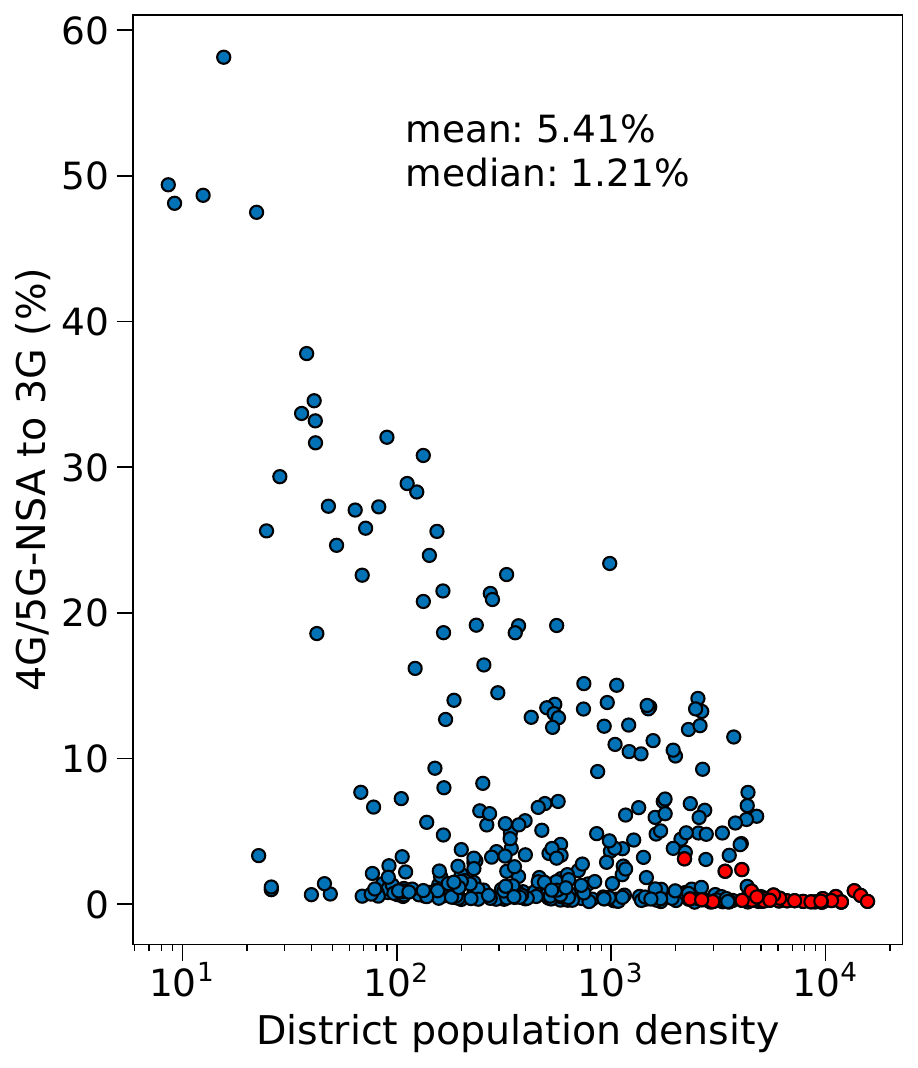}
         % \vspace{-2mm}
         \caption{}
         \label{subfig:heatmaps_rat_to_rat_to3G}
     \end{subfigure}
     \hfil
    \begin{subfigure}[H]{0.25\textwidth}
         \centering
         \includegraphics[width=\textwidth]{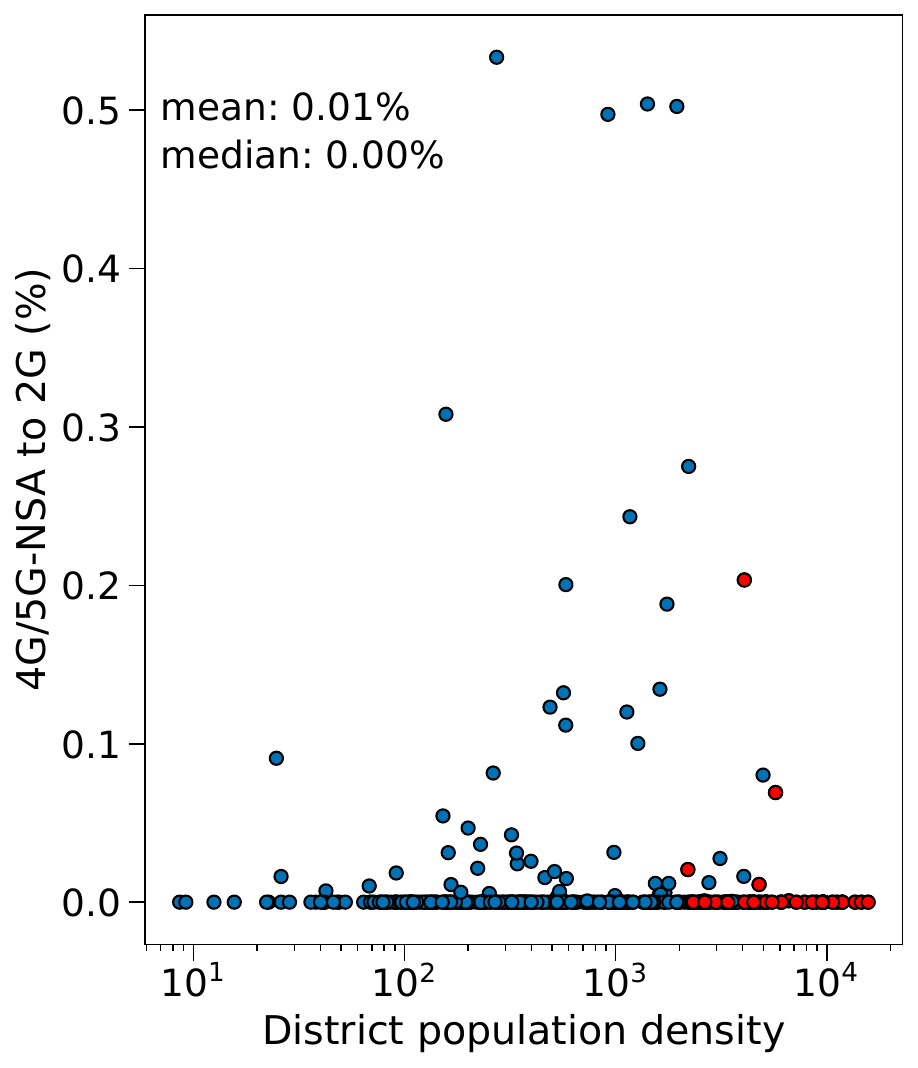}
         % \vspace{-2mm}
         \caption{}
         \label{subfig:heatmaps_rat_to_rat_to2G}
     \end{subfigure}
     % \vspace{-2mm}
     \caption{Distribution of (a) intra 4G/5G-NSA, (b) 4G/5G-NSA to 3G, (c) 4G/5G-NSA to 2G, HOs across districts. Y-axes have different scales.} 
     % \Description{Distribution of (a) intra 4G/5G-NSA, (b) 4G/5G-NSA to 3G, (c) 4G/5G-NSA to 2G, HOs across LADs. Y-axes have different scales.}
     \label{fig:heatmaps_rat_to_rat}
     % \vspace{-1mm}
\end{figure*}

\noindent\textbf{HO Duration.} Figure~\ref{fig:ecdf_ho_time} illustrates the signaling times of HOs (see definition in \S\ref{sec:evaluation-metrics}), revealing that 95\% of intra 4G/5G-NSA HOs complete within $\approx$90ms (median of 43ms). These results align with previous studies~\cite{ho_times_3, ho_times, wheels_2023}. In contrast, HOs from 4G/5G-NSA to 3G are one order of magnitude longer, with a median of 412ms and their 95th percentile exceeding 1s. The latency further increases for vertical HOs to 2G, where the median time matches the 95th percentile for HOs to 3G ($\approx$1s), and the 95th percentile stretches beyond 3.8s. Even if these HO types rarely occur (see Table~\ref{tab:horizontal_vs_vertical_hos}) their large duration reveals a clear negative impact of vertical HOs. We delve into the duration of HOFs in \cref{sec:areas_of_attention} through their causes.

\noindent\textbf{HOs per District.}
Figure~\ref{fig:heatmaps_rat_to_rat} provides a comprehensive view of HO dynamics across \revtwo{districts} in the \revtwo{studied country.} In this way, we are able to pinpoint the areas that are more dependent on newer/older RATs. Notably, \revtwo{densely populated} urban \revtwo{districts} \revtwo{--~which include the districts of the capital city~--} exhibit a high penetration of 4G/5G-NSA (up to 99.92\% of all HOs, see Figure~\ref{subfig:heatmaps_rat_to_rat_intra}), while some less populated rural areas show more transitions to legacy RATs. \revtwo{For example, in the 6\% least densely populated districts, HOs to 3G account for 26.5\% on average of all HOs, and reach up to 58.1\% for a specific remote district (Figure~\ref{subfig:heatmaps_rat_to_rat_to3G}). Likewise, the percentage of transitions to 2G remains marginal for most of the districts, with a maximum of $\approx$0.5\% for 4 specific districts. (Figure~\ref{subfig:heatmaps_rat_to_rat_to2G}).}

\noindent\textbf{\textit{Key takeaways}}: \textit{(i) 94\% of HOs are intra 4G/5G-NSA, and are triggered by smartphones. (ii) HOs to 3G/2G take up to 3.8 seconds (pct-95) to execute and still represent 6\% of all HOs. (iii) \revtwo{The most densely populated urban areas} rely almost exclusively on 4G/5G-NSA for HOs (>99\%); \revtwo{less densely populated} rural areas still use older RATs \revtwo{(HOs to 3G are up to 58.1\% in a remote area and on average 26.5\% in the least densely populated districts)}. This analysis helps the MNO to identify areas where a great volume of 4G and 5G-capable devices are frequently using legacy RATs, thus building a realistic strategy towards their decommissioning.}

\subsection{Mobility across Device Types}
\label{sec:mob_diff_device}

This section examines the relationship between UEs' mobility and their HO performance. We first characterize mobility metrics across different device types. Then, we analyze the relation between the mobility metrics of the UEs and the HOF rate that they experience, serving as an indicator of how these UEs suffer from service disruptions.

\begin{figure}[t]
    \centering
    \begin{subfigure}[H]{0.235\textwidth}
         \centering
         \includegraphics[width=\columnwidth]{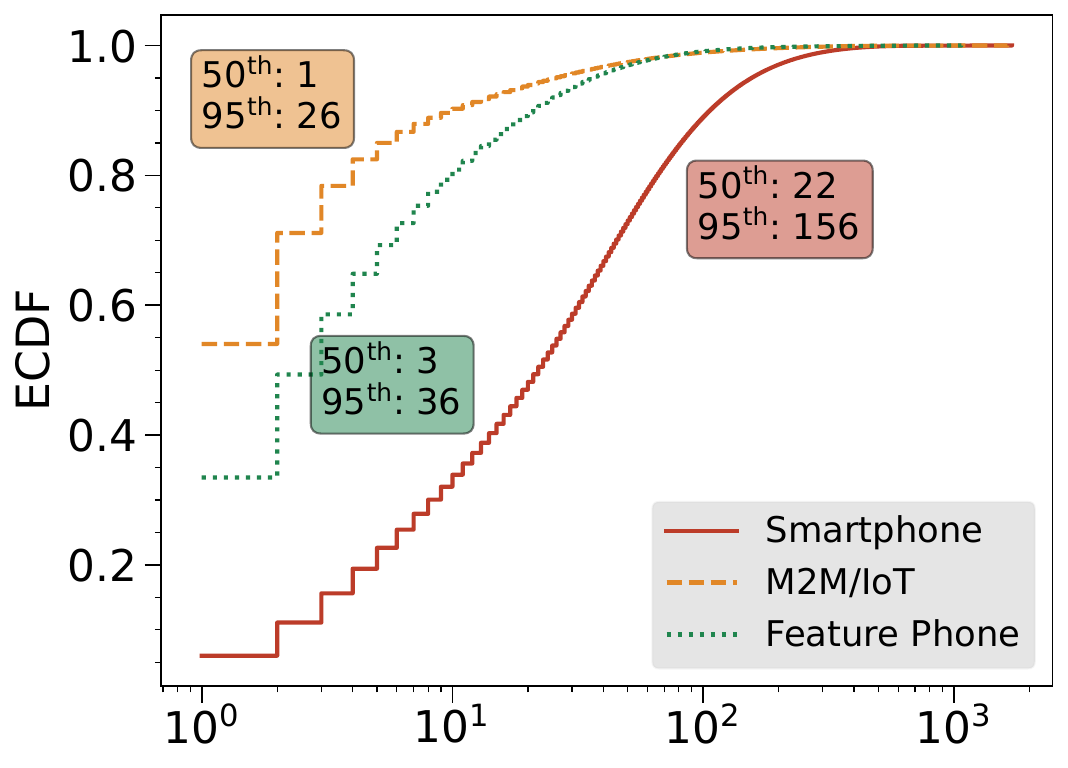}
         \caption{Number of sectors}
         \label{fig:ecdf_num_sectors_device}
     \end{subfigure}
     \begin{subfigure}[H]{0.235\textwidth}
         \centering
         \includegraphics[width=\columnwidth]{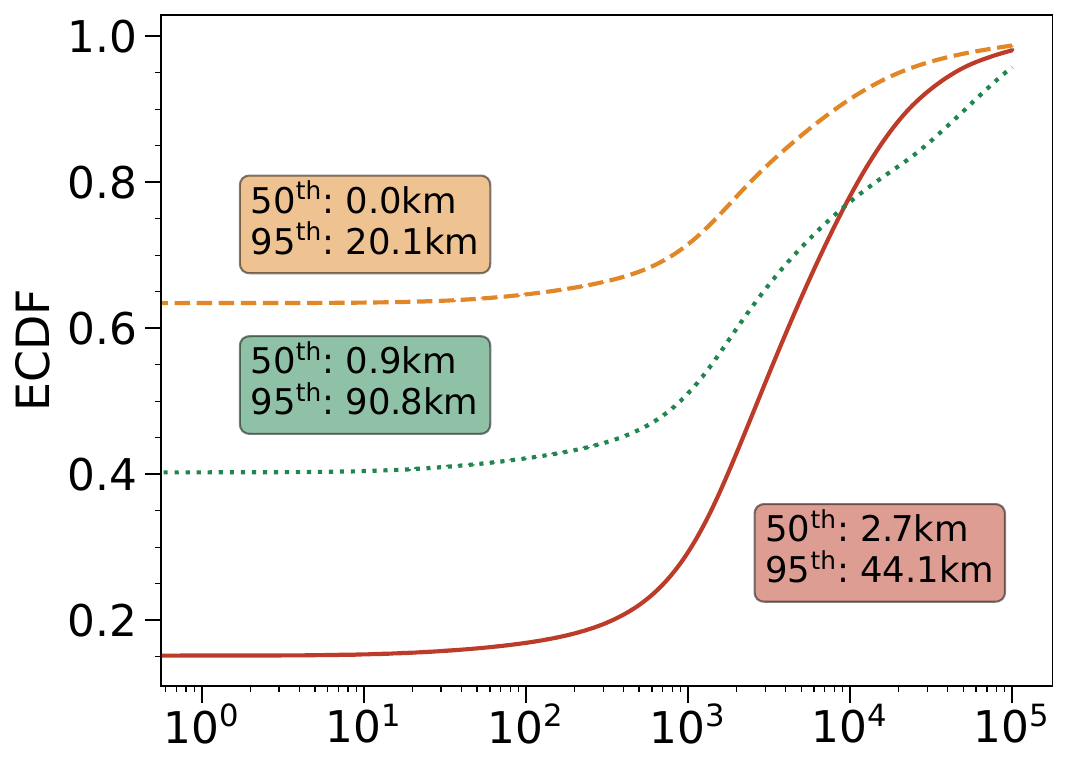}
         \caption{Gyration (m)}
         \label{fig:ecdf_gyration_device}
     \end{subfigure}
     % \vspace{-2mm}
     \caption{Mobility metrics across device types.}
     % \Description{Mobility metrics across device types.}
     \label{fig:ecdf_gyration_num_sectors}
\end{figure}

We take as a reference the two mobility metrics described in \cref{sec:evaluation-metrics}: radius of gyration and number of sectors. Figure~\ref{fig:ecdf_gyration_num_sectors} shows the empirical cumulative distribution function (ECDF) of both mobility metrics across device types. Overall, we observe that smartphones are considerably more mobile than the two other types, exhibiting a median of 22 distinct visited sectors per day, and a median radius of gyration of 2.7km. Conversely, the majority of M2M/IoT devices and low-tier feature phones are more static, with median values of 1 and 3 visited sectors per day, respectively, and a median gyration of 0.0km and 0.9km. This reflects that these UEs are mostly static, and the few HOs that these devices experience are typically between sectors in the same sites. 

Given the heterogeneity of M2M/IoT vertical applications, there are devices in the 95\textsuperscript{th} percentile that show high mobility, with gyrations of 20.1km for M2M/IoT devices (see Figure \ref{fig:ecdf_gyration_device}). These UEs mainly correspond to modems and routers that are deployed in fast-moving vehicles (e.g., trains), integrated into modern cars, embedded in industrial equipment, or wearable IoT devices carried by users who typically travel long distances. While feature phones (green line) surpass smartphones (red line) at around the 80\textsuperscript{th} percentile, the former comprise only about 1\% of the total UEs, while the latter makes up approximately 60\% of UEs (see Figure \ref{subfig:manufacturers}).

\noindent\textbf{Manufacturer Impact.}
\label{subsec:ho_manufacturers}
We assess whether higher HO counts and HOF rate correlate with specific UE manufacturers (e.g., due to a suboptimal mobility management implementation). We observe that the distribution of UEs is remarkably unbalanced across the \revtwo{studied country}, e.g., Samsung and Apple smartphones are considerably more common in densely populated areas. To make a fair comparison and account for potential deviations due to the area itself (e.g., \revtwo{population}, deployment density) --~see Figure~\ref{fig:num_hos_lads}~-- we create a metric that makes a unified comparison of UE manufacturers at the \revtwo{district} level. That is, in each \revtwo{district} we get the average HOs per UE for a specific manufacturer and divide it by the average HOs per UE including all manufacturers within that \revtwo{district} (i.e., \textit{normalized \revtwo{district}-level HO});\footnote{Some manufacturers have few devices in specific \revtwo{districts.} We exclude \revtwo{district}-manufacturer pairs that account for <1k devices.} and similarly for the HOF rate (i.e., \textit{Normalized \revtwo{district}-level HOF rate}). A value greater than 1 indicates that UEs of a specific manufacturer generate more HOs (or HOF rate) on average than the total population of UEs in the same \revtwo{district.}

\begin{figure}[t]
    \centering
    \includegraphics[width=\columnwidth]{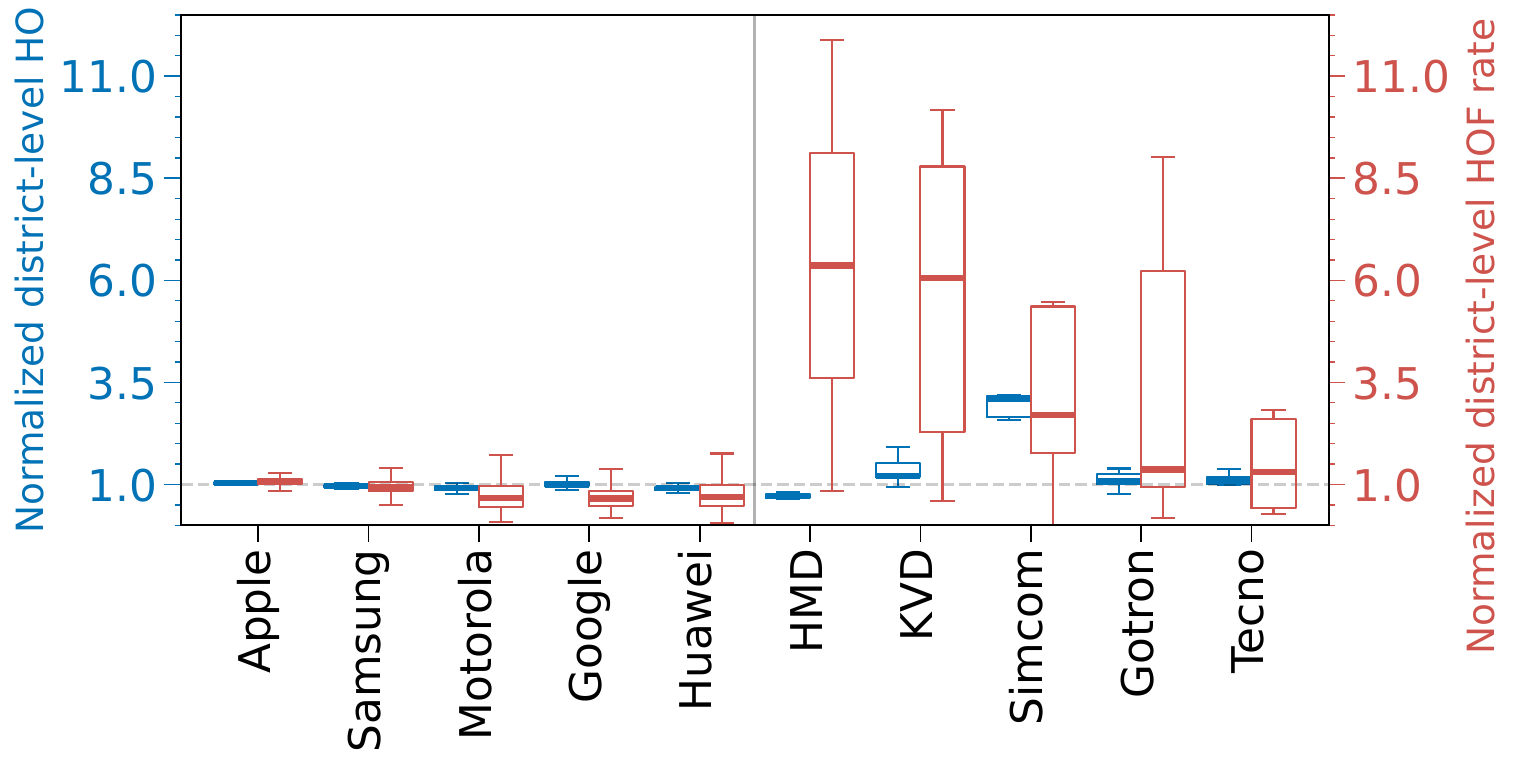}
    % \vspace{-5mm}
    \caption{Normalized \revtwo{district}-level HOs (left) and HOFs (right) per UE manufacturer. Boxplots include top-5 UE manufacturers and the top-5 UE manufacturers with greater median HOF values.}
    % \Description{Normalized district-level HOs (left) and HOFs (right) per UE manufacturer. Boxplots include top-5 UE manufacturers and the top-5 UE manufacturers with greater median HOF values.}
     \label{fig:ho_hof_peers}
     % \vspace{-4mm}
\end{figure}

Figure \ref{fig:ho_hof_peers} shows the results for the top-5 manufacturers in the \revtwo{studied country} (see Figure~\ref{subfig:manufacturers}), as well as the 5 manufacturers exhibiting the highest \textit{Normalized \revtwo{district}-level HOF rate}, based on the median behavior across all \revtwo{districts} (see boxplots). 
For the top-5 manufacturers ratios are close to 1, which means that devices behave similarly to their peers in the same \revtwo{district}, both in terms of HOs and HOF rate. Specifically, we observe that Apple smartphones, the most popular ones ($\approx$32\% of all UEs), generate slightly more HOs per UE and HOF rates than other devices (respectively +4\% HOs and +8\% HOF w.r.t. their peers). Likewise, Google smartphones are the ones that experience the smallest HOF rates (-27\% w.r.t. their peers). Moreover, we find that some manufacturers show high HOF rates, such as KVD smartphones or HMD feature phones (+600\% HOF rate), as well as others that generate higher HOs per UE, such as Simcom M2M/IoT (+293\% HOs per UE). 

\noindent\textbf{\textit{Key takeaways}}: \textit{(i) Different UE types exhibit different mobility patterns; smartphones are, on median, the ones connecting to more distinct sectors (22 sectors per day), with a daily median radius of gyration of 2.7km. (ii) The most popular device manufacturers exhibit a consistent behavior in terms of HOs ($\pm$10\% of variation between them). While HOF rates are considerably small, some manufacturers (e.g., Google) exhibit lower HOF rates (-27\%) than other manufacturers. For some niche manufacturers, we find high HOF rates (up to +600\%) and HO counts (up to +293\%). Based on these results, we conjecture that manufacturer-specific mobility management implementations and application-specific usage correlate with HO performance.}

\section{Handover Failure Analysis}
\label{sec:areas_of_attention}

This section provides an in-depth analysis of HOFs. Initially, we examine the daily patterns of HOFs and their correlation with key mobility metrics. Next, we explore the causes of HOFs from the network's perspective and present modeling techniques that assess how network features at the radio sector level influence the HOF rate. Our analysis puts the spotlight on the need to reduce the network's complexity by decommissioning legacy RATs.

\subsection{HOF Patterns \& Impact }\label{subsec:HOF-patterns}

\begin{figure}[t]
    \centering
    \includegraphics[width=\columnwidth]{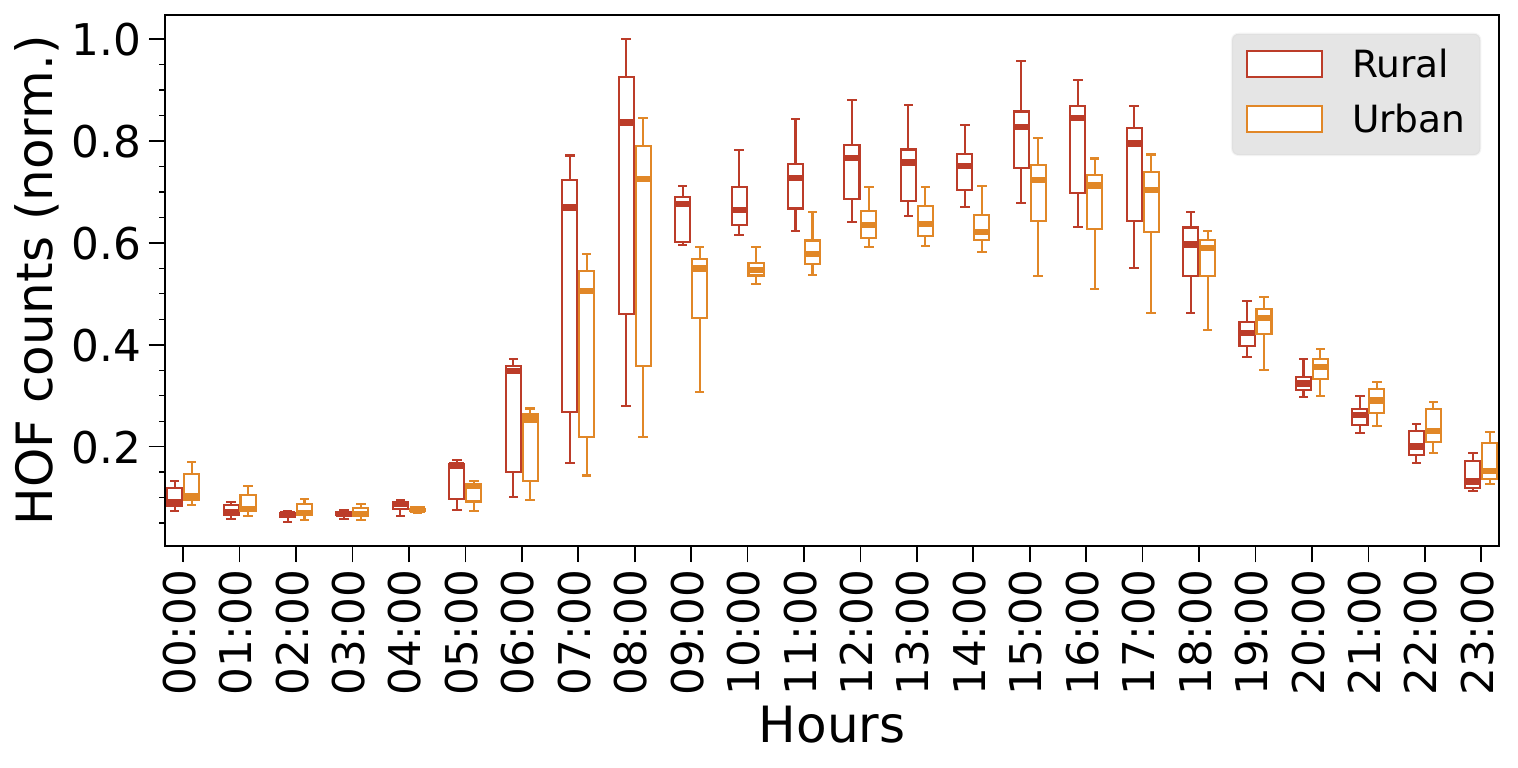}
    \vspace{-0.5cm}
    \caption{HOF counts per hour in urban and rural areas, normalized separately with the number of active sectors in each
    \revtwo{class} (i.e., urban/rural).}
    % \Description{HOF counts per hour in urban and rural areas, normalized separately with the number of active sectors in each setting (i.e., urban/rural).}
     \label{fig:hof_hourly}
     % \vspace{-4mm}
\end{figure}

\noindent\textbf{HOF Patterns.} 
We analyze the temporal evolution of HOF rate (see \S\ref{sec:evaluation-metrics}) along the day, aggregating data over the 4-week period. Figure~\ref{fig:hof_hourly} presents the hourly evolution of HOFs, where boxplots aggregate data from all active radio sectors at a specific hour. To comply with the privacy policies of the MNO and account for the different distribution of sectors in rural and urban areas, we have separately normalized the hourly HOFs for rural (urban) areas with the number of active sectors observed in rural (urban) settings (see Figure \ref{fig:ho_and_sectors_per_day}, bottom). Overall, we observe that \revtwo{HOFs reach a local peak during the morning commuting time [7:00--9:00), and a lower local peak can be observed during the afternoon commuting time [15:00--18:00). Moreover,} urban areas experience fewer HOFs compared to rural ones, especially during peak hours\revtwo{; e.g.,} the median HOF count is 32.4\% higher in rural areas than in urban ones during [7:00--8:00). We conjecture that this pattern is likely due to the more limited 4G/5G coverage in these areas, which makes 4G and 5G-capable devices fall back more frequently on older RATs (i.e., 2G, 3G) to keep connectivity. We further delve into this aspect in \cref{sec:regression} by modeling the negative impact of vertical HOs on HOFs and inspecting the causes of such failures.

\begin{figure}[t]
    \centering
    \begin{subfigure}[H]{0.235\textwidth}
         \centering
         \includegraphics[width=\columnwidth]{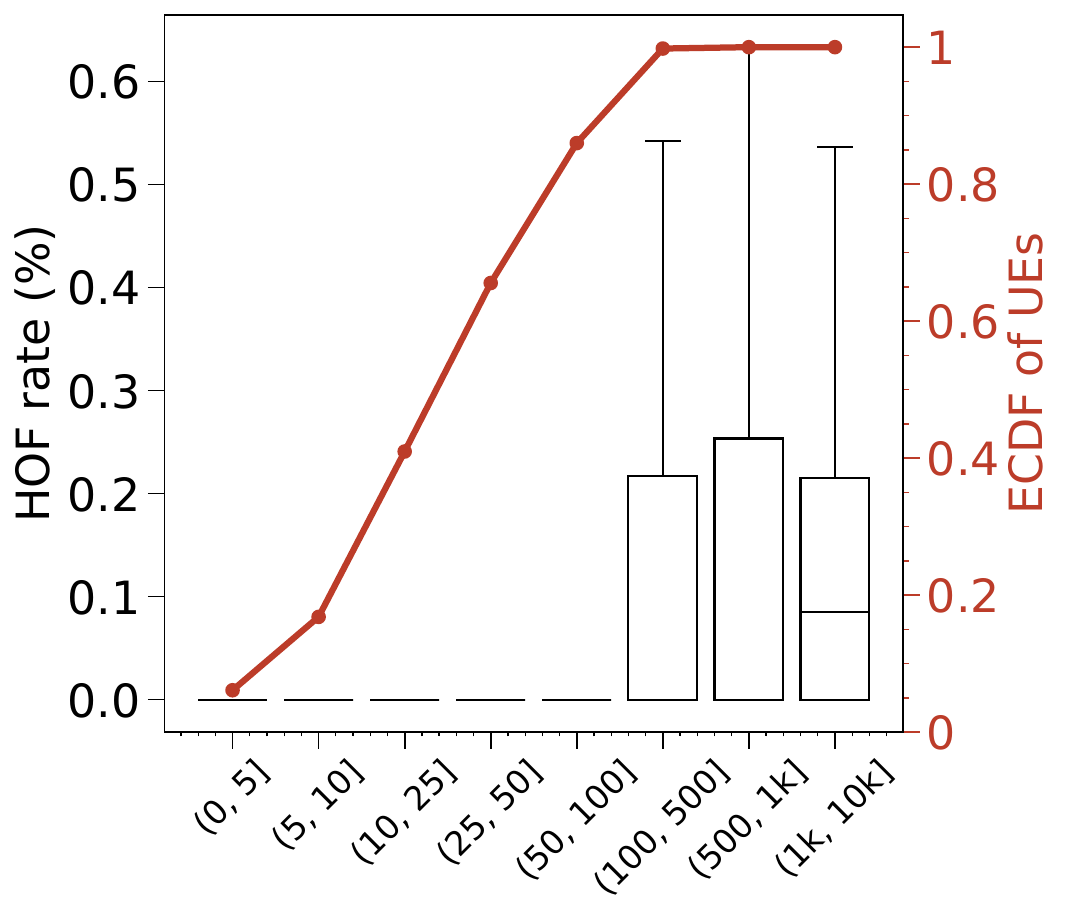}
         \caption{Number of sectors.}
         \label{fig:ecdf_num_sectors_hof}
     \end{subfigure}
     \begin{subfigure}[H]{0.235\textwidth}
         \centering
         \includegraphics[width=\columnwidth]{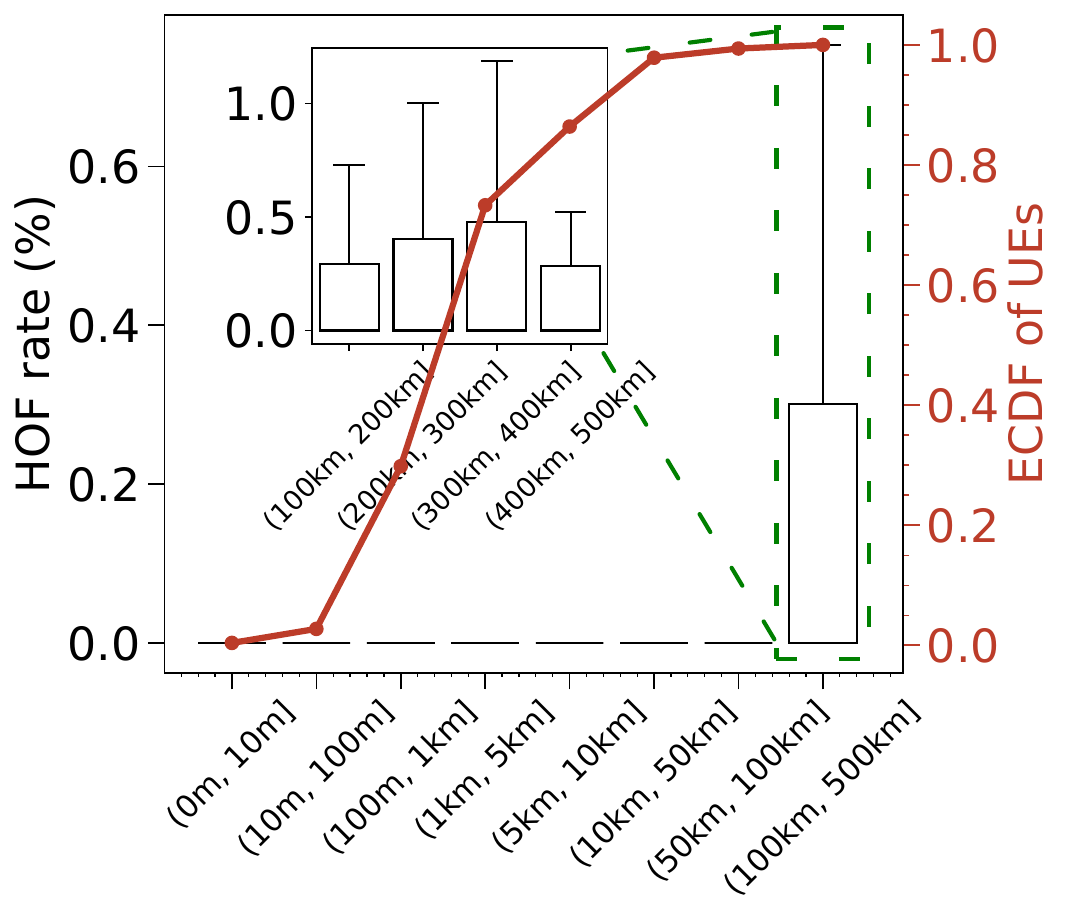}
         \caption{Gyration}
         \label{fig:ecdf_gyration_hof}
     \end{subfigure}
     \caption{HOF rate (left y-axis) and ECDF for the number of UEs (right y-axis) w.r.t. binned device-level mobility metrics (log scale).}
     % \Description{HOF rate (left y-axis) and ECDF for the number of UEs (right y-axis) w.r.t. binned device-level mobility metrics.}
     \label{fig:ecdf_hof_gyration_num_sectors}
     % \vspace{-4mm}
\end{figure}

\noindent\textbf{HOFs \& Mobility.}
We explore the association of radius of gyration and number of sectors with the HOF rate. In Figure~\ref{fig:ecdf_hof_gyration_num_sectors}, the left y-axis shows the daily average HOF rate for the UEs according to the number of sectors (Figure~\ref{fig:ecdf_num_sectors_hof}), or radius of gyration (Figure~\ref{fig:ecdf_gyration_hof}). Meanwhile, the right y-axis displays the ECDF for the number of UEs along the bins in the x-axis \revtwo{(in log scale)}.

Concretely, Figure~\ref{fig:ecdf_num_sectors_hof} shows that the HOF rate is close to zero for 87\% of the UEs, which connect to 100 or less sectors per day. For the remaining 13\% of the UEs (traveling $>$100 sectors), the HOF rate slightly increases (up to 0.4\% for pct-75), but the median is still close to zero; except for $<$0.0001\% of the UEs that connect to $>$1k sectors and have a median HOF rate of 0.1\%. Similarly, from Figure~\ref{fig:ecdf_gyration_hof}, HOFs mainly occur in devices that move within a radius higher than 100km (which is the case for 0.007\% of the devices, see the right y-axis), with the HOF rate reaching up to 0.4\% (pct-75). Yet, the median HOF rates remain close to zero for all bins. \rev{We observe that the devices with increased mobility ($>$100 visited sectors, $>$100km radius of gyration) are mostly smart/feature-phones (90\%) and M2M devices (10\%) -- such as modems, routers and IoT wearables -- attached or carried in fast-moving vehicles, like trains. It is interesting to note that UEs with $<$10km radius of gyration and $<$50 visited sectors, which show almost zero HOF rate, include a very similar share of UE types (85\% smart/feature-phones and 15\% M2M); which confirms that the increase in HOFs in UEs with higher mobility metrics cannot be explained by an unequal distribution of UE types in this group.}

\noindent\textbf{\textit{Key takeaways}}: \textit{ (i) Rural areas (with sparser deployments) suffer from 32.4\% more HOFs during peak hours than urban. (ii) A small number of UEs with high mobility metrics daily ($>$100 visited sectors, $>$100km radius of gyration) experience a non-negligible HOF rate (0.4\% for pct-75); the number of visited sectors and the radius of gyration are good predictors to flag UEs that can potentially experience high HOF rates.}

\subsection{Causes of HOFs}
\label{subsec:causes_hof}

\begin{figure}[t]
    \centering
    \begin{subfigure}[H]{0.235\textwidth}
         \centering
         \includegraphics[width=\textwidth]{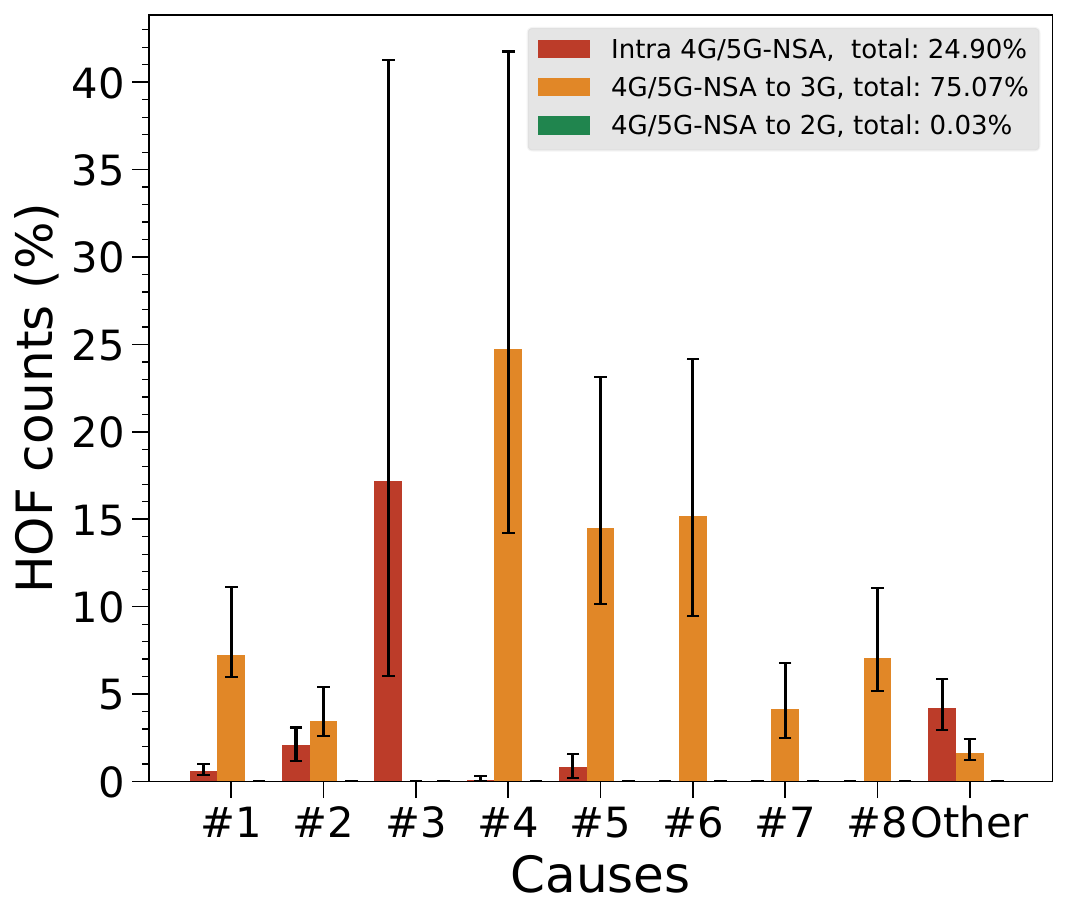}
         \caption{}
     \end{subfigure}
     \begin{subfigure}[H]{0.235\textwidth}
         \centering
         \includegraphics[width=\textwidth]{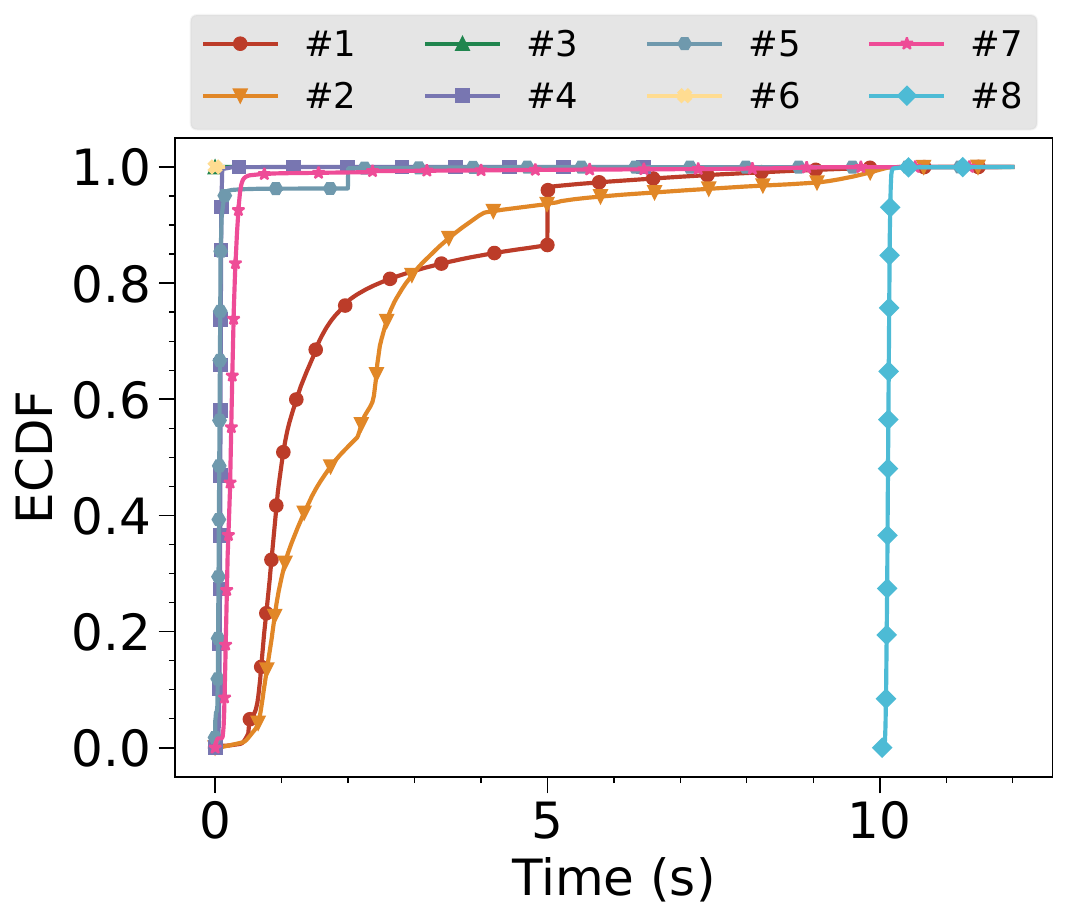}
         \caption{}
     \end{subfigure}
     \caption{(a) Percentage of HOF causes w.r.t. the total HOFs; (b) Distribution of HO signaling time per cause.}
     % \Description{(a) Percentage of HOF causes w.r.t. the total HOFs; (b) HO signaling time that each cause requires.}
     \label{fig:causes_rat_ecdf}
     % \vspace{-6mm}
\end{figure}

\begin{figure}[t]
    \centering
    \begin{subfigure}[H]{0.45\textwidth}
         \centering        \includegraphics[width=\columnwidth]{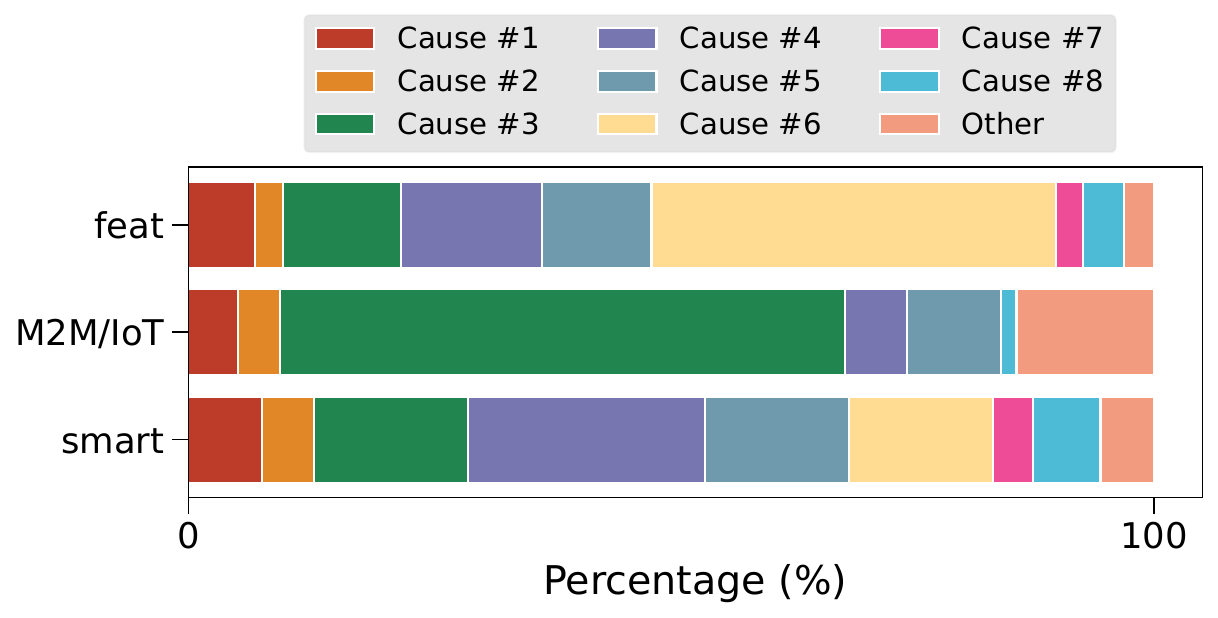}
         \caption{\rev{Causes vs device types.}} \label{subfig:causes_vs_class}
     \end{subfigure}
     \begin{subfigure}[H]{0.45\textwidth}
         \centering
         \includegraphics[width=\columnwidth]{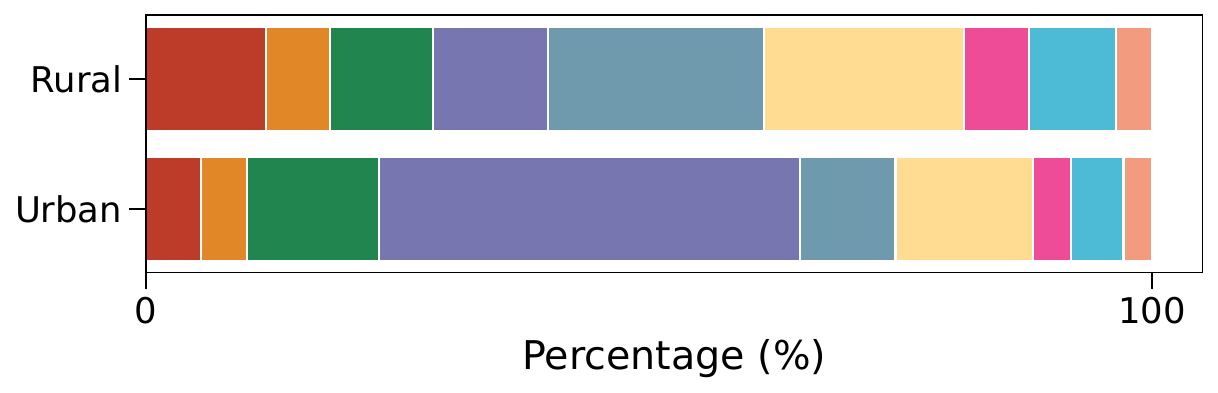}
         \caption{\rev{Causes vs urban/rural areas.}} \label{subfig:causes_vs_ru11ind2}
     \end{subfigure}
          \begin{subfigure}[H]{0.45\textwidth}
         \centering
         \includegraphics[width=\columnwidth]{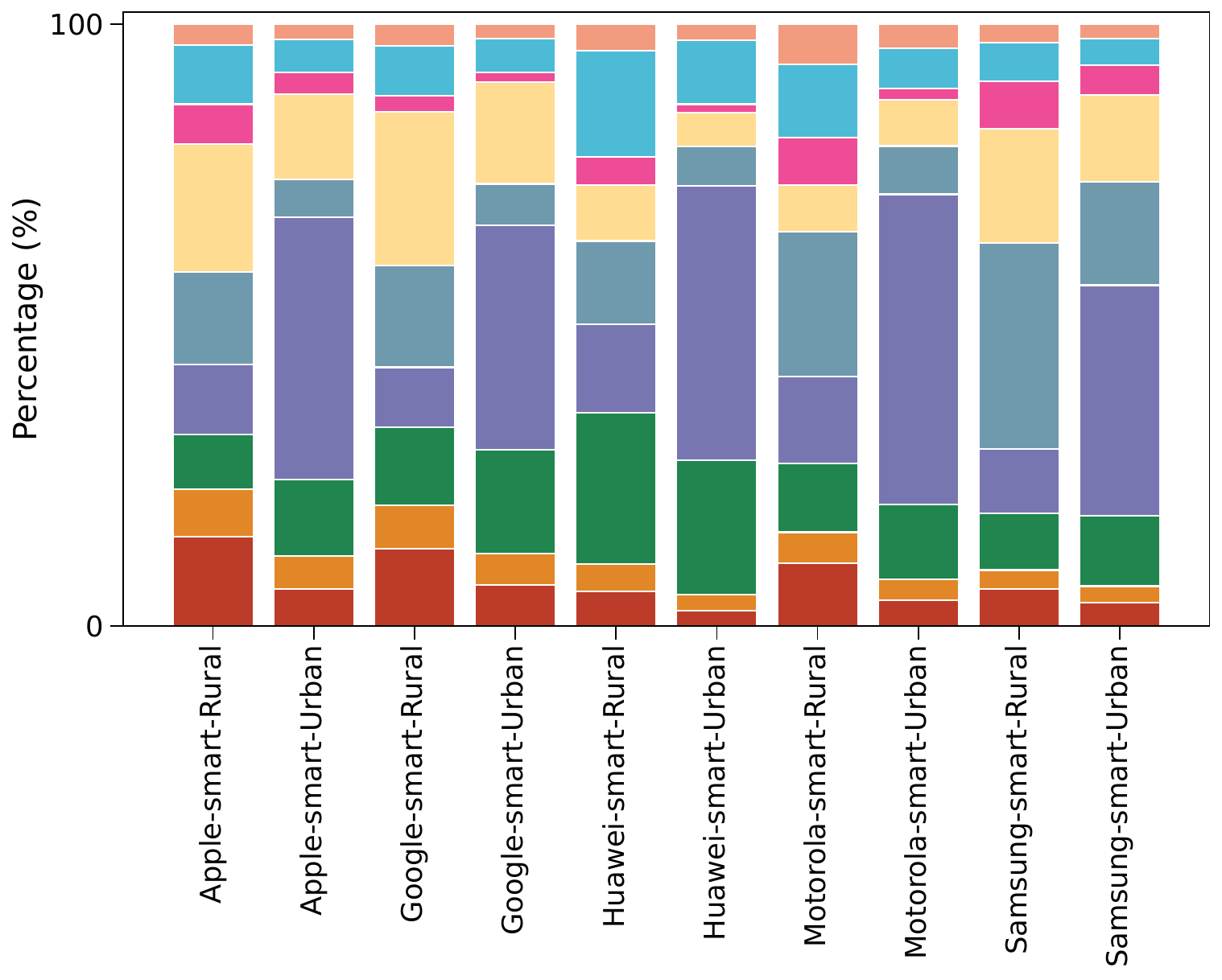}
         \caption{\rev{Causes vs top-5 smartphone manufacturers.}} \label{subfig:causes_vs_manuf}
     \end{subfigure}
     \caption{\rev{Stacked bars showing the percentages of HOF causes (each bar adds to 100\%).}}
     % \Description{Percentages of HOF causes.}
     \label{fig:extra_causes}
     % \vspace{-5mm}
\end{figure}

We study the HO failures using cause codes delineated by the 3GPP standards \cite{3gpp_36_413, 3gpp_29_274} and the antenna vendors. In total, we collect 1k+ different causes for the failures. Our causes analysis complements prior studies that had exclusively focused on the user side, being mostly coarser, and solely for specific devices and failure types \cite{lte_rails_2019, li2021nationwide}. In Figure \ref{fig:causes_rat_ecdf}a, we present the HOF counts in percentage, by calculating the HOF for each cause and dividing it by the total HOFs per day. We also plot alongside the minimum and maximum values observed in this period (i.e., 4 weeks). Our analysis reveals that ($i$)~92\% of all HOFs occur because of 8 causes from the 1k+ that exist, and that ($ii$) 75\% of all HOFs occur in transitions from 4G/5G-NSA to 3G, with the remainder (i.e., $\approx$25\%) associated with intra 4G/5G-NSA HOs. HOFs attributable to transitions to 2G represent 0.03\% of all. This distribution highlights the real-world implications of managing a layered cellular deployment, where $<\!6\%$ of handovers are vertical handovers to 3G, and the remaining $95\%$ are intra 4G/5G-NSA handovers. We present next the 8 most common handover failure causes. Additional insights for the reasons for HOFs in rural/urban areas, different smartphone manufacturers, and UE types can be seen in \rev{Figure~\ref{fig:extra_causes}}.

\noindent$\bullet$ \textbf{Cause \#1}: ``The source sector canceled the HO'' relates to the cancellation of an ongoing or prepared handover. HO Cancellation procedure \cite{3gpp_36_413} can occur for several reasons, such as timeouts on the MSC, cell site, or issues with the size of the Forward Relocation Request \cite{3gpp_29_274}. This cause is predominantly observed in HOs to 3G, affecting 7.3\% to 11.2\% of cases daily, which is significantly higher compared to intra 4G/5G-NSA and 4G/5G-NSA to 2G HOs ( $\!<1\%$ per day). We observe that this failure cause affects evenly all UE types, but is 50\% more prevalent in rural than in urban areas (see \rev{Figure~\ref{fig:extra_causes}}).

\noindent$\bullet$ \textbf{Cause \#2}: ``The signaling procedure was aborted due to interfering S1AP Initial UE Message \cite{3gpp_36_413}''. This error involves the interruption of the signaling process by an initial message to the MME, which includes critical user information and service requests. This issue affects 2\% of intra 4G/5G-NSA HOs and 3.4\% of HOs to 3G, but not HOs to 2G. 

\noindent$\bullet$ \textbf{Cause \#3}: ``Signaling procedure was rejected due to invalid target sector ID'' occurs when the target sector ID is not recognized or if there are configuration issues with the MME pool area (i.e., a collection of MMEs configured to serve any common part of a radio network). This is the main reason for failure in intra 4G/5G-NSA HOs, accounting for an average of 17.2\% of the failures, and reaching up to 41.3\%. From this cause, 59\% of M2M/IoT devices fail (see \rev{Figure~\ref{fig:extra_causes}}).

\noindent$\bullet$ \textbf{Cause \#4}: ``Load on target sector is too high'' indicates that the target sector cannot accommodate the HO due to resource constraints. It is the most common reason for failure in HOs to 3G (up to 42.3\% of all HOFs), affecting 25\% of the failures per day, on average. It happens mainly during peak hours in dense urban areas (see Figure \ref{fig:ho_and_sectors_per_day}), causing 42\% of the total HOFs there (see \rev{Figure~\ref{fig:extra_causes}}).

\noindent$\bullet$ \textbf{Cause \#5}: ``MME detects a HO-related failure in the target MME, SGW, PGW, cell, or system''; these types of infrastructure-related outages occur for 14--23\% of HOs to 3G, and for 0.8--1.6\% of intra 4G/5G-NSA HOs. This cause does not pinpoint precisely the reason that the HOF occurred; however, it is important to note that this is the extent of information that is available to the MNO.

Causes \#6, \#7 \#8 are specific to HOs from 4G/5G-NSA to 3G. We provide more information in the sequel.

\noindent$\bullet$ \textbf{Cause \#6}: ``The Single Radio Voice Call Continuity (SRVCC) service is not subscribed by the UE'' affects 15.2\% of HOs to 3G on average, peaking at 24.1\%. SRVCC is a scheme used with VoLTE (Voice over LTE) and ensures seamless handovers of voice calls from packet-switched (PS), like 4G, to circuit-switched (CS) networks, like 2G and 3G \cite{3gpp_23_216, 3gpp_23_008}. We note that this failure occurs primarily in rural areas and in feature phones, where the MNO still relies mostly on 3G to ensure the support of voice services (see \rev{Figure~\ref{fig:extra_causes}}).

\noindent$\bullet$ \textbf{Cause \#7}: Like Cause \#6, Cause \#7 is associated with SRVCC HOs, and it occurs when ``the MSC responds with PS to CS Response with cause indicating failure'' during SRVCC HO preparation; it affects about 4.2\% of all HOs \cite{3gpp_23_216}. We note that it affects almost no M2M/IoT device and occurs twice as often in rural than in urban areas (see \rev{Figure~\ref{fig:extra_causes}}).

\noindent$\bullet$ \textbf{Cause \#8}: ``No Forward Relocation Complete or Notification was received before the max time for waiting for the relocation completion expires'', affecting 7.1\% of HOs. Forward Relocation Complete message is sent to the source MME/SGSN to indicate the HO has been successful. We observe this cause $\times 3$ in M2M/IoT devices w.r.t. smartphones and feature phones (see \rev{Figure~\ref{fig:extra_causes}}).

\noindent\textbf{HOF Duration per Cause.} Figure \ref{fig:causes_rat_ecdf}b complements the HOF analysis by delineating the HO duration associated with the 8 causes of failure. Causes \#3 and \#6 result in failures that prevent the initiation of the HO (i.e., signaling time equals 0ms), with the former attributed to an invalid target sector ID and the latter to SRVCC service not being permitted for the UE. Cause \#4, linked to insufficient resources in the target sector, exhibits a median duration of 81ms and a 95th percentile of 97ms. More prolonged delays in HO signaling are caused by Causes \#1 and \#2, where the HO is halted due to cancellation by the source or interference, respectively, leading to medians of 1--2s and 95th percentiles of 5--6s. Notably, Cause \#8, associated with timeout failures, demands the most extended signaling, with a median $>$10s and 95\% of cases occurring in $<\!10.2$s. Our study aligns with existing works, such as \cite{lte_rails_2019, polycorn_trains23}, in demonstrating the increased duration involved in HOFs.

\noindent\textbf{\textit{Key takeaways}}: \textit{(i) Despite the prevalence of 4G and 5G, 75\% of all failures concern HOs from 4G/5G-NSA to 3G, and 25\% of them happen due to high load \revtwo{in} the target sector. \rev{(ii) 59\% of M2M/IoT UEs and 42\% of feature phones fail due to Cause \#3 and Cause \#6, respectively. (iii) 42\% of HOFs in urban areas occur due to Cause \#4, while Causes \#5 and \#6 account for 20\% each, in rural areas.} (iv) The duration of HOs that fail due to timeouts (Cause \#8) or cancellations (Cause \#1) exceeds on average 2s (for the former case it reaches 10s). These numbers highlight the noticeable
outage duration caused by HOFs in the network. }

\subsection{Modeling HOFs} 
\label{sec:regression}

We aim to understand which sector-level features contribute to HOFs, \rev{by isolating and combining the effect of various parameters and ensuring our findings in the previous sections are robust against potential biases or unaccounted variables}. We reorganize the dataset using as dependent variable the \emph{daily} HOF rate of each source sector, and use as covariates the sector-level features in Table \ref{tab:regression-features}. This creates 6.7M observations. Our hypothesis is that the HO type is the primary factor influencing HOF rates. Note that while HOs to 3G amount to only 5.86\% of the total HOs \revtwo{(\cref{ho-types})}, they are responsible for 75\% of all HOFs \revtwo{(\cref{subsec:causes_hof})}.

\begin{table}[!t]
{\small{ \centering
     \caption{Regression covariates.}
     \vspace{-2mm}
    \begin{tabular}{p{3.8cm}p{3.83cm}}
    \toprule
    Feature & Values \\    
    \midrule
    Number of HOs per day & $\geq 0$ \\
    RATs & 4G/5G-NSA, 3G, 2G \\   
    \revtwo{District population} & $\geq 0$ \\
    Sector \revtwo{Region} & West, South, North, \revtwo{Capital area}\\
    Area \revtwo{Type} & Rural / Urban \\    
    Antenna Vendor & 4 vendors (V1, V2, V3, V4)\\ 
    \bottomrule
    \end{tabular}
    \label{tab:regression-features}}}
    \vspace{-2mm}
\end{table}

A first look into the data demonstrates that HOs to 2G and 3G are associated with substantially higher failure rates with medians of 21.42\% and 5.85\% respectively, compared to 0.04\% for HOs to 4G/5G-NSA; and this disparity persists even when we focus on the failed HOs and filter outliers, see Appendix~\ref{sec:appendix-regression}. We further perform an analysis of variance (ANOVA) test \cite{anova} (log-transforming the HOF rates) which verifies the significance of this effect ($p<0.001$); and the same conclusion is reached using the Kruskal-Wallis test~\cite{kruskal_wallis}. We repeat these tests, with the same findings, even when controlling for variations in the area and antenna vendor.

\begin{table}[!t]
{\small{ \centering
     \caption{Linear model coefficients for $\log(HOF\,rate)$.}
     \vspace{-2mm}
    \begin{tabular}{p{2.5cm}p{1.2cm}p{1.8cm}p{1.3cm}}
    \toprule
    Feature & Coef. & 95\% CI & P-value \\    
    \midrule
    Intra 4G/5G-NSA & $-2.77$ & -2.77, -2.76 & $0$\\    
    4G/5G-NSA$\rightarrow$3G & $5.12$  & 5.12, 5.13 & $0$\\
    4G/5G-NSA$\rightarrow$2G & $6.82$  & 6.76, 6.88 & $0$ \\
    \bottomrule
    \end{tabular}
    \label{tab:regression-coefficients}}}
    % \vspace{-2mm}
\end{table}

Accordingly, we use a generalized linear regression model (with log transformation) to quantify the effect of RAT on HOFs. We first run a univariate model to facilitate interpretation. We find that HOs to 3G (2G) increase the HOF rate by 166\% (915\%, respectively) compared to HOs to 4G/5G-NSA, see Table \ref{tab:regression-coefficients}. We repeat this analysis while controlling for the other covariates and filtering the outliers (HOF rate<50\%, number of HOs per day in $[50,30k]$), finding the same result with slightly smaller intensity (coefficients of 5.48 and 4.77 instead of 6.82 and 5.12), as can be seen in Table~\ref{tab:regression-linear-all-features}. From the remaining covariates, the antenna vendor has a significant but smaller effect, which we also verify with an ANOVA test. These findings are also robust to alternative models (step-wise covariate selection and removing HOs to 2G), including also a quantile linear regression model.
The details of these additional tests are deferred to Appendix \ref{sec:appendix-regression}.

\noindent\textbf{\textit{Key takeaways}}: \textit{By modeling HOFs and investigating different covariates (see Table \ref{tab:regression-features}), we verify our hypothesis that the HO type is the main factor shaping the observed HOF rates: HOs to 3G (2G) increase the HOF rate by 166\% (915\%, respectively) compared to HOs to 4G/5G-NSA.}

\section{Related Work}\label{sec:related_work}

In terms of measurement approaches, the vast majority of studies rely on (mainly rooted) UEs and collect traces from their cellular modems \cite{mob_support_2018, lte_rails_2019, raca_4G_dataset, xiao_4G_dataset, li2020beyond, 5G_ues_2020, 5G_wild_2021, 5G_mmwave_2022, raca_5G_dataset, vivisecting_2022, wheels_2023, li2021nationwide}. For instance, \cite{mob_support_2018} and \cite{li2020beyond} build their mobility analysis upon Mobile Insight \cite{mobileinsight_mobicom16} with rooted phones, while \cite{raca_4G_dataset, raca_5G_dataset} use the G-NetTrack Pro monitoring tool \cite{G-NetTrack-Pro}. These solutions are confined to certain chipset manufacturers and have limited data collection granularity (orders of seconds, instead of msec as in the current study). Other works study mobility patterns in one \cite{china_20} or a few cities \cite{5G_wild_2021}, such as Minneapolis \cite{5G_ues_2020, lumos5G}, Chicago \cite{5G_ues_2020, 5G_mmwave_2022}, Atlanta \cite{5G_ues_2020}, and Rome~\cite{rome22}. These studies provide valuable information, yet their spatial focus does not facilitate insights across larger scales (e.g., countrywide) and in varied settings (e.g., urban/rural areas). The works of \cite{xiao_4G_dataset, lte_rails_2019, polycorn_trains23} conduct extensive 4G performance measurements on high-speed rails in China, and \cite{vivisecting_2022, wheels_2023} study mobility management policies in 4G/5G. The collected data in these relevant works are related to certain mobility patterns, and a subset of users, and do not contain network-side data.

\begin{table}[!t]
\rev{
{\footnotesize{ 
\centering
\caption{{\small{\rev{Regression Summary: Linear Model, All Covariates.}}}}
\begin{tabular}{p{2.65cm} p{1.2cm} c c c}
\toprule
Feature & Coeff. & Std Err & t value & Pr($>|t|$) \\    
\midrule
(Intercept) & $-3.10$ & $0.0217$ & $-143$ & $0$ \\   
HO type: 4G/5G-NSA$\rightarrow$2G & $5.48$ & $0.118$ & $46.4$ & $0$ \\           
HO type: 4G/5G-NSA$\rightarrow$3G & $4.77$ & $0.00150$ & $3169$ & $0$ \\ 
Number of daily HOs & $-2.84 \cdot10^{-5}$ & $0$ & $-331$ & $0$ \\    
\revtwo{Area Type}: Rural & $0.260$ & $0.00272$ & $95.5$ & $0$ \\            
\revtwo{Area Type}: Urban & $0.190$ & $0.00258$ & $73.4$ & $0$ \\
\revtwo{Antenna} Vendor: V2 & $0.115$ & $0.00173 $ & $66.7$ & $0$ \\
\revtwo{Antenna} Vendor: V3 & $0.719$ & $0.0203$ & $35.3 $ & $0$ \\         
\revtwo{Antenna} Vendor: V4 & $0.0629$ & $0.0222$ & $2.84$ & $0.49$ \\      Sector Region: North & $-0.0728$ & $0.0216$ & $-3.57$  & $4.05\cdot10^{-6}$ \\ Sector Region: South & $-0.0168 $ & $0.00166$ & $-10.1$ & $2.28\cdot10^{-6}$\\ 
Sector Region: West & $0.398$ & $0.0204$ & $19.5$ & $3.89\cdot10^{-66}$ \\   
\revtwo{District population} & $-1.75\cdot10^{-7}$ & $0$ & $-61.6$ & $0$ \\             
\midrule
\multicolumn{5}{l}{$N=3857074$, \quad RMSE=$1.023$, \quad $R^2=0.8265$, \quad	AIC=11121590}  \\
\bottomrule
\label{tab:regression-linear-all-features}
\end{tabular}}}}
% \vspace{-4mm}
\end{table}

Our study, on the other hand, records \emph{all} mobility events from a commercial MNO network with $\approx$40M UEs connected, with msec granularity, during 4 weeks, and for the entire \revtwo{territory of a European country}; it is not limited to specific routes, cities, mobility modes, or user types. To date, only a few works study HOs from the operator's perspective, as this involves technical challenges~\cite{mob_support_2018} and requires in-network measurements (see Figure \ref{fig:HO-archi}). Namely, \cite{feher_2012} suggests an approach to categorize and minimize undesired Ping-Pong (PP)\footnote{PP HO occurs when a UE is handovered from a source to a target sector, and then back to the source, under a short, predefined time.} HOs based on a restricted dataset with 1.7k UEs; and \cite{pp_ho_2023} investigates PP HOs using 13 days of data from a network operator in a Mediterranean area. Our study differs from these works due to the scale, coverage, and granularity of measurements (all active connections of a \revtwo{top-tier MNO at the country level}; see Table \ref{tab:statistics}), and due to the fusion of different datasets (about UEs and population) that allows drawing fresh insights, e.g., about the impact of HOs and HOFs on different RATs, device types, \rev{and areas (rural/urban, and \revtwo{district level})}.

Specifically, in terms of measurement results, our findings about the HO duration are on par with previous studies, e.g. \cite{wheels_2023, vivisecting_2022, lte_rails_2019}, and provide additional insights, e.g., about the effect of RAT, finding that inter-RAT HOs are the most impactful. Several studies measured the volume of HOs \cite{yuan2022understanding, 5G_ues_2020, 5G_wild_2021}, finding, e.g., horizontal HOs to be more frequent in 5G-SA and 4G and vertical HOs in 5G-NSA~\cite{yuan2022understanding}. Here, we enrich these results by dissecting the HOs per RAT and UE manufacturer/type, analyzing their temporal pattern over 4 weeks, and their \revtwo{relation to the demographic distribution} over the \revtwo{studied country}, with district granularity (\revtwo{300+ districts}), thus refining the typical urban/rural categorization of prior studies~\cite{pp_ho_2023}.

Furthermore, leveraging our unique network-side dataset, we \revtwo{characterize} HOFs (cause \emph{and} duration) using detailed antenna vendor-specific information. Prior studies, inhibited by their UE-side data, have mainly studied the effect of user speed on HOFs~\cite{li2020beyond} or used coarser categorization, e.g., 2 possible causes~\cite{lte_rails_2019}, or analyzed general connectivity failures for specific devices~\cite{li2021nationwide}. Given that HOs were found to affect significantly the user-perceived network performance, our work can inform the design of new HO policies, such as~\cite{yuan2022understanding, icellspeed_2020, cellbricks_2021}, and guide the optimization of network deployment and RAT upgrades.

\section{Discussion}\label{sec:limitations}

\textbf{Limitations.} 
\revthree{Datasets and actual, unnormalized, numbers in the figures cannot be published openly due to privacy guidelines of the MNO (see Appendix \ref{sec:appendix-ethics}).} \revthree{Moreover,} at the time we conducted our study, the 5G-SA deployment of the MNO was still in its early stages, with a limited range of (mostly test) UEs actively using it. Thus, we focus on 5G-NSA, which relies on the 4G EPC for mobility management. In other words, we cannot explicitly capture the HOs to/from 5G radio sectors, since the EPC only sees their corresponding 4G radio sector anchor. In addition, the studied HOs have 4G/5G-NSA as the source RAT, and 4G/5G-NSA, 3G, or 2G as the target. In other words, apart from the horizontal HOs in 4G/5G-NSA, we focus on the specifics of how/when/why users downgrade to older RATs, and not the other way around, \rev{given that users spend more than 82\% of their time and 94.5\% of their traffic in 4G/5G (see \cref{sec:first_look_at_network})}.

\revthree{Lastly,} ($i$)~for this study we did not have access to HO configuration parameters and policies, which are dynamically configured by proprietary solutions from equipment vendors (see \cref{sec:mobility_management}), \revthree{and ($ii$)~our analysis on \textit{HOFs \& Mobility} is limited to the use of mobility metrics (number of sectors and radius of gyrations) at daily intervals, which may hide correlations that occur at finer time scales. While this paper represents a first attempt to provide an overview on a countrywide scale, we stress the importance for the community to conduct further studies on the previous aspects in order to contrast the coarse-grained correlations found in this study with specific analyses focused on establishing causal relations between various metrics, e.g., by analyzing performance degradation for specific users at the session-level just before and after a HOF occurs.
}

\noindent\textbf{Guidelines\rev{, Implications \& Future Works.}} The progression towards RATs beyond 4G is pivotal for realizing greater network speeds, minimizing latency, and improving performance. \rev{However, the findings of this study underline the significant challenges that older RATs present in HO performance, which directly impacts the users' Quality of Experience.} Although \revthree{some of} these technologies are close to their sunset, they \rev{continue to play a crucial} role in many operational networks (as the studied one) and hence, require attention to maintain users' satisfaction. It is crucial for network operators to monitor and report activity in the legacy RATs,  so as to design realistic strategies towards fully decommissioning them. \rev{The gradual phasing out of older RATs should be carefully managed to avoid any unintended negative consequences on network reliability and user experience. This may involve transitions of certain regions or user segments to newer RATs before others, based on usage patterns and network demands.}

\rev{The implications of these findings extend to HO policies, which need to be revisited in light of the persistent issues associated with older RATs (see \cref{sec:areas_of_attention}).} \revthree{Network operators should consider adopting more dynamic and adaptive HO algorithms (e.g., in response to failures and mobility patterns) that can handle the specific challenges posed by these legacy technologies. These solutions should be tailored to the various causes of HOFs associated with HOs to 3G/2G and intra 4G/5G-NSA HOs (see \cref{subsec:causes_hof}), and therefore, to the different time granularity in which these HOs occur (i.e., hundreds/thousands of msecs and tens of msecs, respectively, see \cref{ho-types}).}

Further investigation into the role of device manufacturers and operating systems is also essential. Optimizing network performance is not solely the responsibility of network operators; additional \emph{coordinated} studies that examine the internal mechanisms of devices and their influence on HOs are required. This entails a deeper understanding of how manufacturers and operating systems interact with network procedures and the identification of HO improvement opportunities (e.g., where HO durations are prolonged)\rev{.}

\revthree{Future work could also explore the impact of HOFs on performance metrics, such as throughput, voice/data accessibility, and success rates \cite{chroma}, from the operator’s perspective. This would enable us to better understand the relation between HO performance and users' Quality of Service (QoS). 

Additionally, large-scale analyses like the one presented here often face challenges in handling, storing, and processing vast datasets, underscoring the need for further research into efficient data sampling techniques.}

\rev{\noindent\textbf{Handover Challenges in 5G and beyond.} While 5G capabilities are expansive, the co-existence of multiple RATs presents a significant challenge, particularly in the context of HO management~\cite{survey_ho, vivisecting_2022}}. \revthree{The integration of 5G with legacy systems like 4G, 3G, or 2G, introduces significant complexities in the HO process, due to the amplified differences in terms of latency, bandwidth, and signaling requirements in 5G, the new mobility features introduced (e.g., dual connectivity \cite{dual_connectivity18}), and the wide range of device types and new services to be supported (e.g., IoT verticals, time-critical communications). For instance, in EN-DC (EUTRA-NR Dual Connectivity) used in 5G-NSA \cite{3gpp_37_340}, two simultaneous connections are established (a 4G master node, and a 5G secondary node) for data plane messages, but only 4G is used in the control plane. This mechanism makes remarkably more complex the HO procedure, as additional signaling messages need to be exchanged due to the presence of the secondary node, resulting in increased time complexity that could be amplified, e.g., in case of PP HOs. To mitigate these challenges, it is imperative to implement differentiated HO policies tailored to the wide range of device types and services supported by 5G; e.g., IoT and time-critical communications require distinct HO strategies to meet specific service level agreements.}

\section{Conclusion}\label{sec:conclusion}

This work provides the first comprehensive, countrywide, analysis of HOs, leveraging data from a leading MNO in \revtwo{a European country,} by studying $\approx$40M users over four weeks. Our findings highlight the critical impact of spatio-temporal factors, RATs, device types, and manufacturers on horizontal and vertical HOs and HOFs, specifying the reasons for the latter, and modeling them with statistical methods. These findings are crucial for understanding and developing new HO mechanisms and policies, and identifying groups of UEs and areas that require enhanced support. In this way, our analysis lays the groundwork for future improvements in network performance, ensuring that the promise of 5G and subsequent generations of cellular technologies can be fully realized.
\begin{acks} 

\rev{We thank the anonymous shepherd and reviewers for their valuable feedback. 
This work has been supported by ($i$) the Spanish Ministry of Economic Affairs and Digital Transformation and the European Union – NextGeneration EU, in the framework of the Recovery Plan, Transformation and Resilience (PRTR) through the UNICO I+D 5G SORUS project, and UNICO I+D 5G 2021 refs. number TSI-063000-2021-3, TSI-063000-2021-38, and TSI-063000-2021-52, ($ii$)  the National Growth Fund through the Dutch 6G flagship project “Future Network Services”, and ($iii$) the European Commission through Grant No. 101017109 (DAEMON) and Grant No. 101139270 (ORIGAMI).}
\end{acks}

\newpage
\bibliographystyle{ACM-Reference-Format}

\bibliography{sec11_references}

\appendix

\section{Ethics}\label{sec:appendix-ethics}
The datasets we leverage in our research are protected under Non-Disclosure Agreements (NDAs) that explicitly forbid the dissemination of information to unauthorized parties and public repositories. The procedures for data collection and storage within the network's infrastructure strictly follow the guidelines set forth by the MNO, and are in full compliance with local regulations. Moreover, while some metrics are computed on the user-level, our data-handling processes strictly focus on generating aggregated, anonymized insights, \rev{without access to the exact locations/trajectories of the users.} No personal and/or contract information was available for this study and none of the authors of this paper participated in the extraction and/or encryption of the raw data. Ultimately, our datasets and research do not involve risks for the mobile subscribers, while they provide new knowledge about the dynamics of mobility management and handovers.

\section{Regression Analysis Details}\label{sec:appendix-regression}

Here, we complement the main regression models presented in \cref{sec:regression} with additional models, which have comparable performance in terms of Root Mean Squared Error (RMSE) and Mean Absolute Error (MAE) with Random Forest (RF) \cite{random_forest}. The results are aligned and support the reported findings. We remind the reader that the analysis is performed on a dataset that records the daily percentage of failed HOs (i.e., HOF rate) during the studied 4-week interval.

\begin{figure}[!t]
    \centering
    \begin{subfigure}[H]{0.45\textwidth}
         \centering
        \includegraphics[width=0.85\columnwidth]{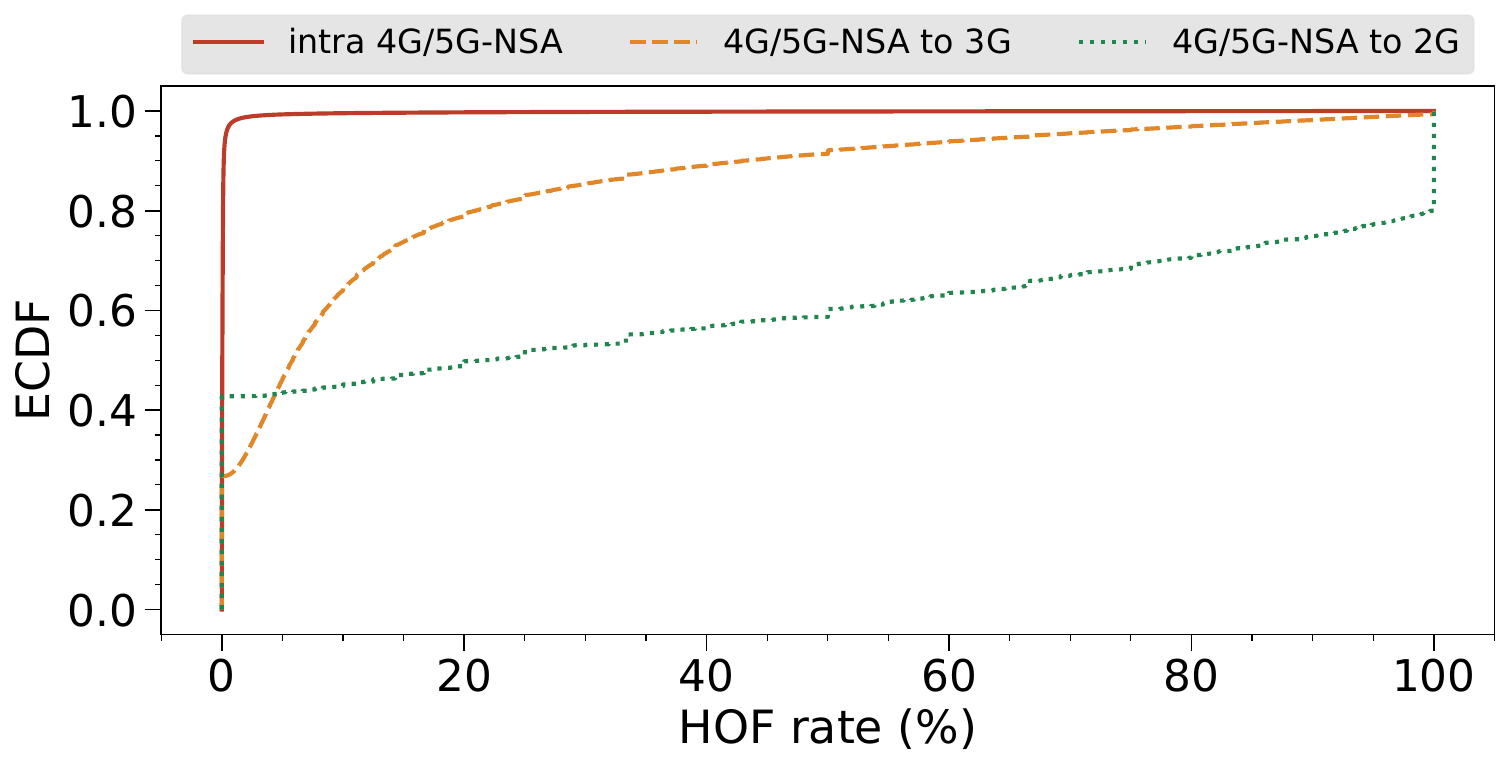}
     \end{subfigure}
     \begin{subfigure}[H]{0.45\textwidth}
         \centering
         \includegraphics[width=0.85\columnwidth]{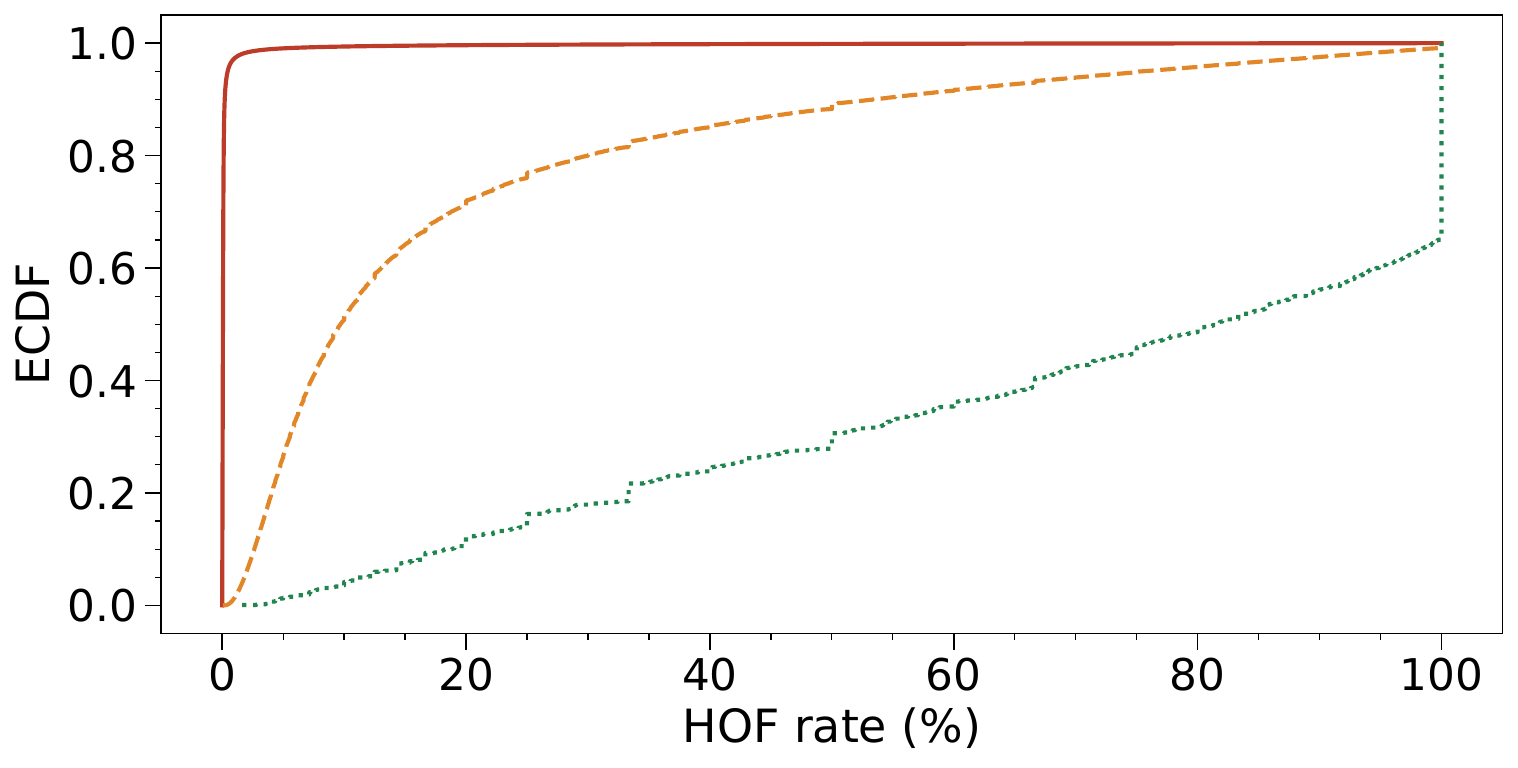}
     \end{subfigure}
     \begin{subfigure}[H]{0.45\textwidth}
         \centering
         \includegraphics[width=0.85\columnwidth]{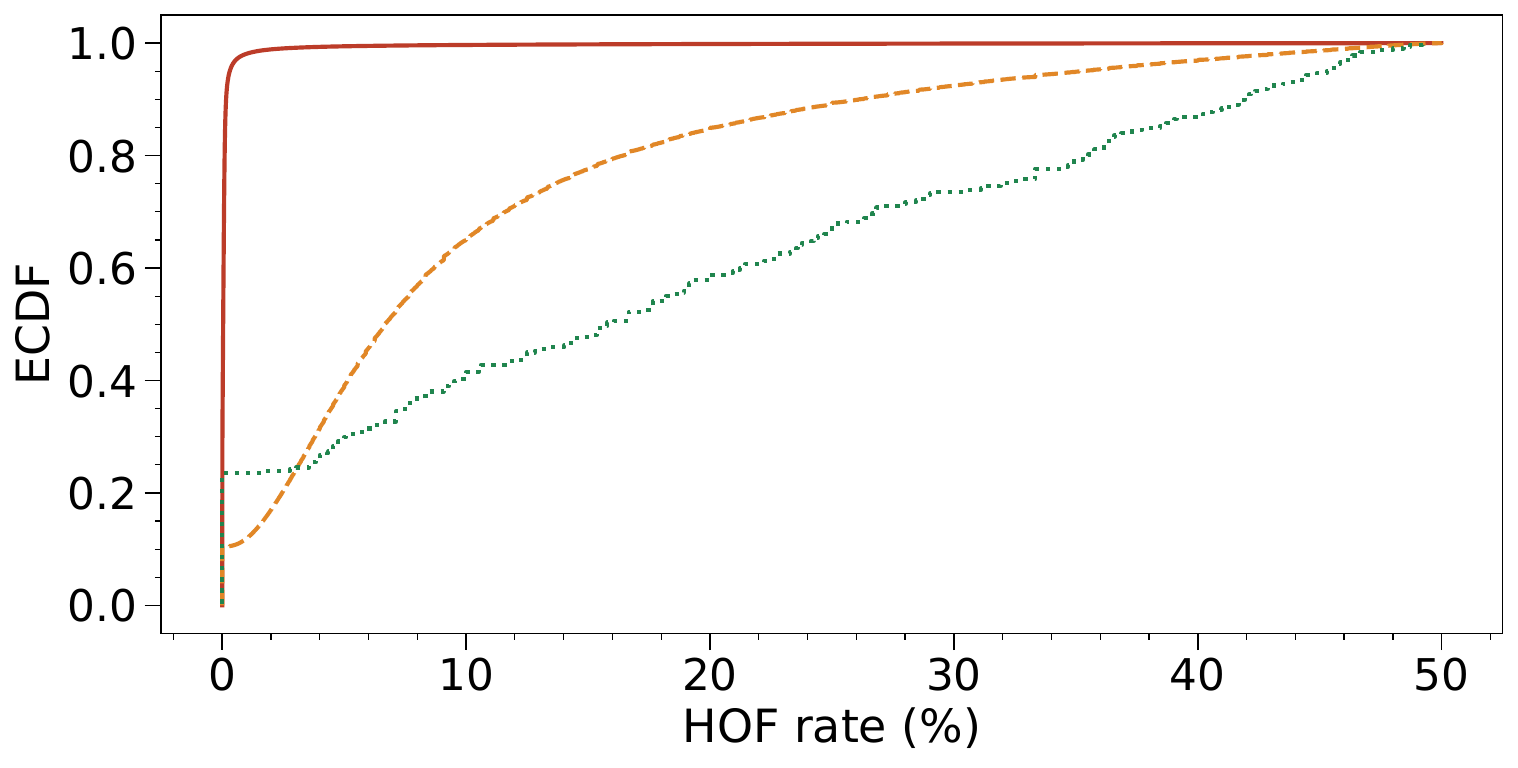}
     \end{subfigure}
     \caption{ECDF of HOF rates for HO type: all HOF rates (top); non-zero HOF rates (middle), HOF rates without outliers (bottom).}
     % \Description{ECDF of HOF rates for HO type: all HOF rates (top); non-zero HOF rates (middle), HOF rates without outliers (bottom).}
     \label{fig:ecdf-HOF-Type}
\end{figure}

We start by plotting the main statistics (boxplots with mean and median values) for the effect of HO type, antenna vendor, and sector area on the HOF rates. We also plot the ECDFs for the first two cases in Figure \ref{fig:ecdf-HOF-Type}, while the summary statistics can be seen in Table \ref{tab:summary-stats}. Performing a one-way ANOVA test we find that the effect of HO type on HOF rate is statistically significant and large ($F(2, 3857071) = 8.01\cdot10^{6}$, $p\!<\!.001$; $\eta^2 = 0.81, 95\%\text{CI \,} [0.82, 1.00]$), and Post-hoc pairwise comparisons (Tukey's HSD) verify that this effect is significant for all HO types. A Kruskal-Wallis test also supports this hypothesis ($p\!=\!0$).

\begin{figure}[!t]
    \centering
    \begin{subfigure}[H]{0.45\textwidth}
         \centering
        \includegraphics[width=0.85\columnwidth]{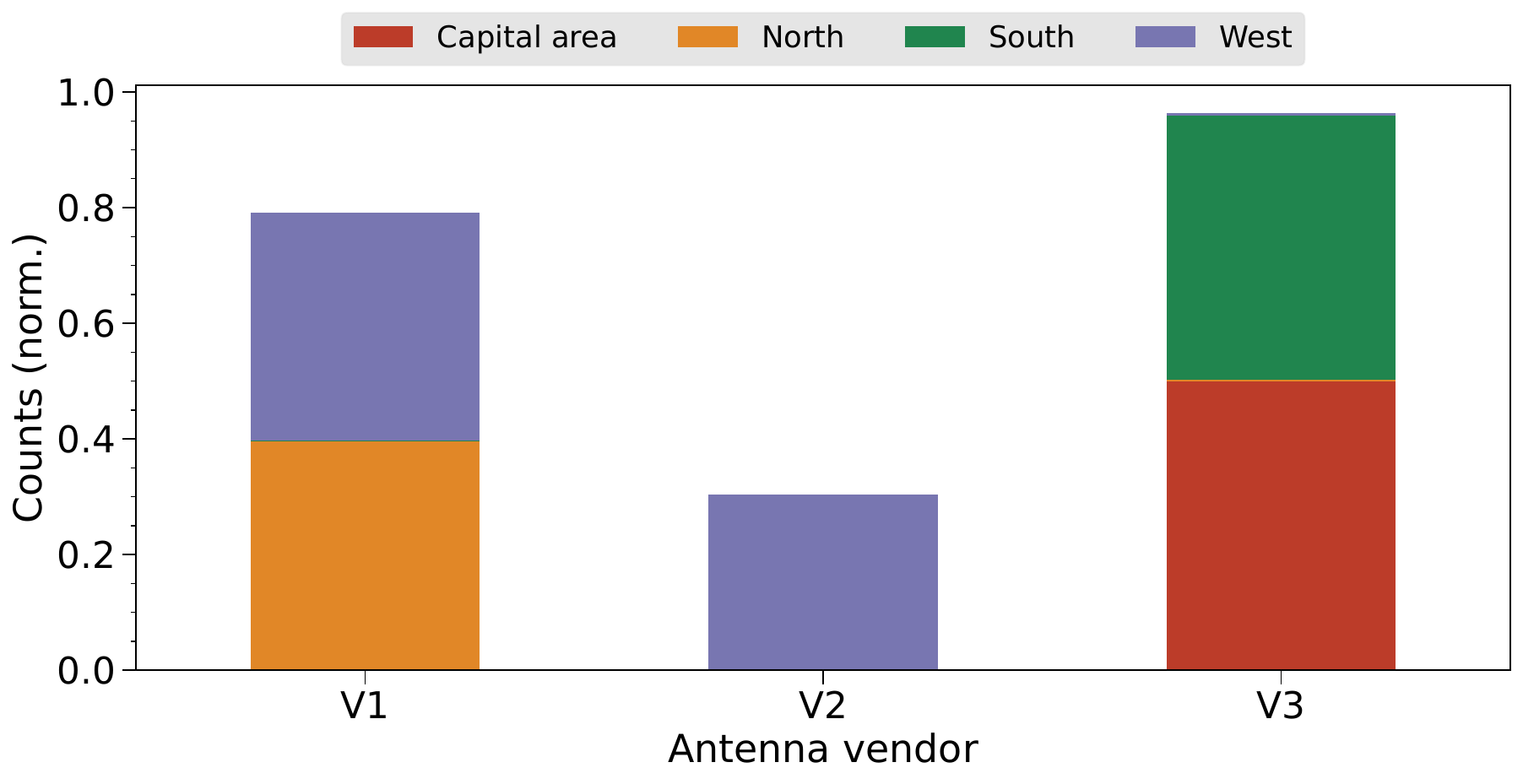}
     \end{subfigure}
     \begin{subfigure}[H]{0.45\textwidth}
         \centering
         \includegraphics[width=0.85\columnwidth]{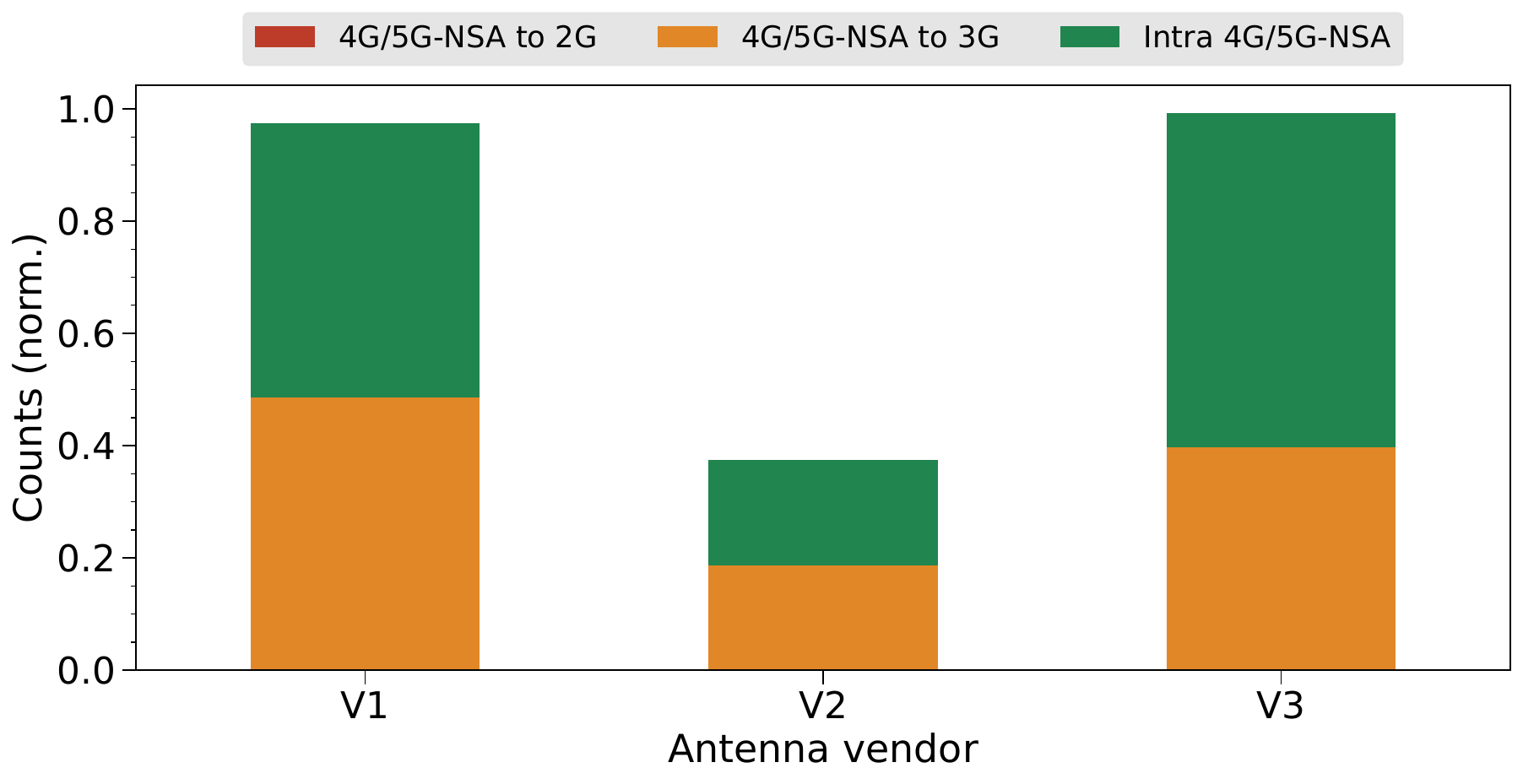}
     \end{subfigure}
     \caption{Antenna vendor per region (top); per  HO type (bottom).}
     % \Description{Antenna vendor per region (top); per  HO type (bottom).}
     \label{fig:HOF-Vendor-Stats}
\end{figure}

\begin{figure}[!t]
    \centering
    \begin{subfigure}[H]{0.45\textwidth}
         \centering
        \includegraphics[width=0.85\columnwidth]{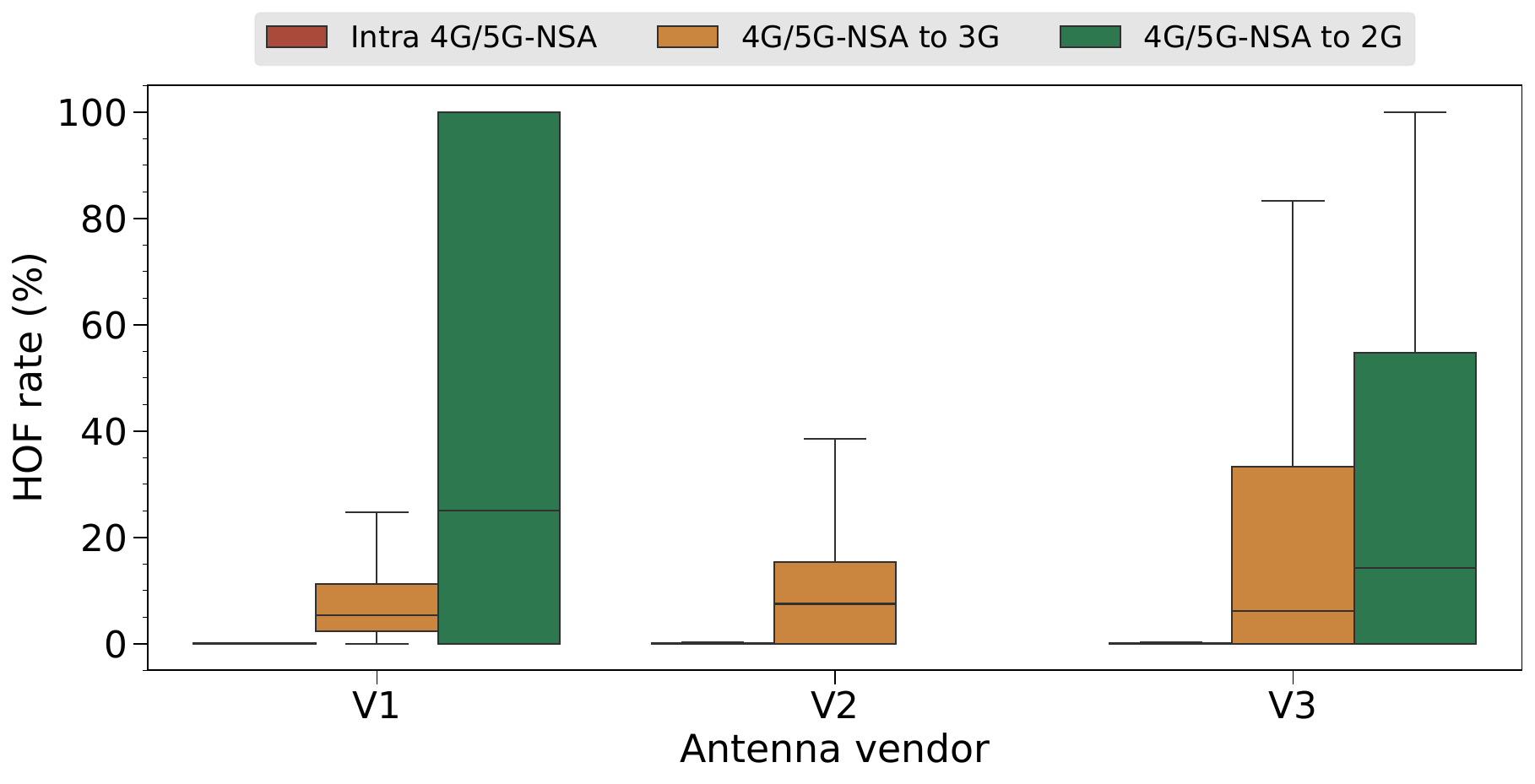}
     \end{subfigure}
     \begin{subfigure}[H]{0.45\textwidth}
         \centering
         \includegraphics[width=0.85\columnwidth]{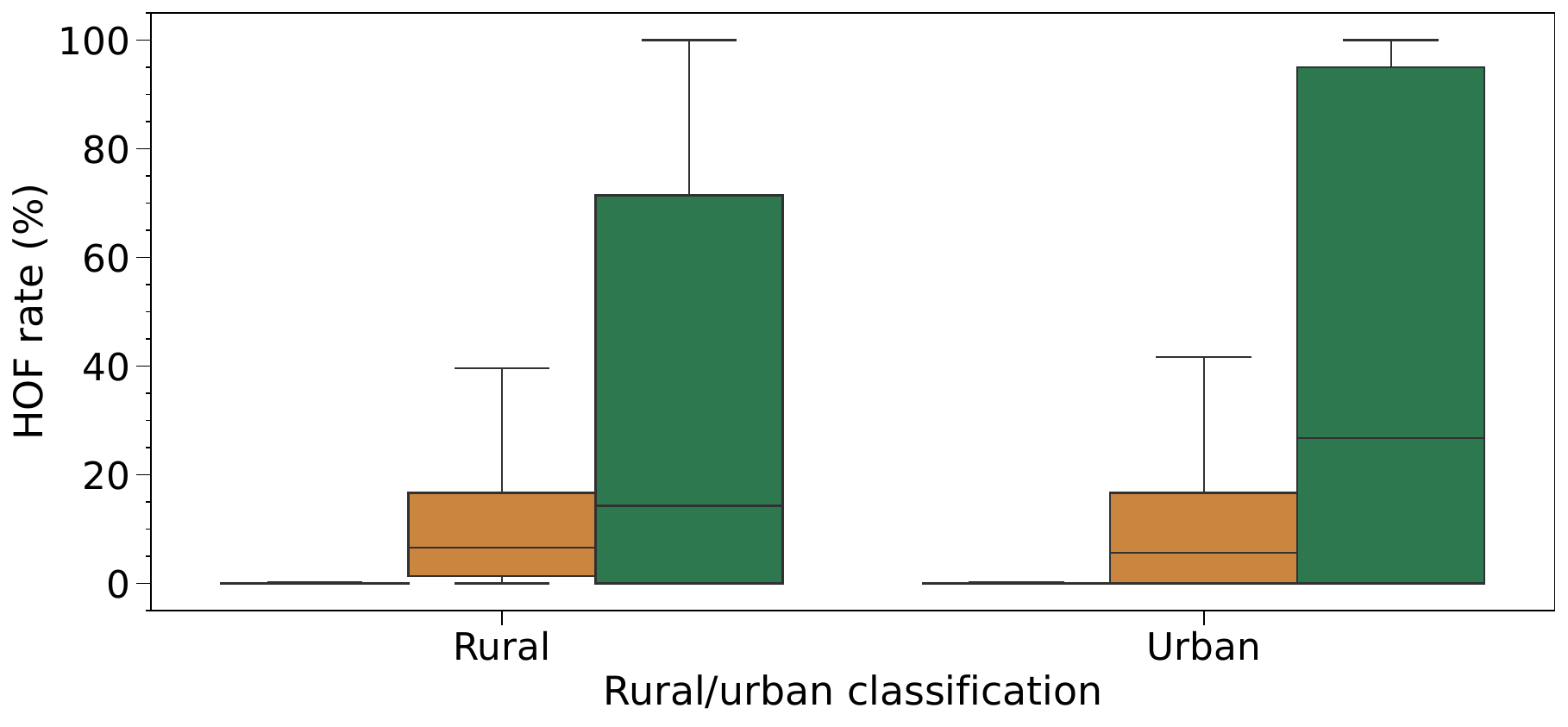}
     \end{subfigure}
     \caption{Boxplots of HOF rates vs antenna vendor (top); vs urban/rural areas (bottom).}
     % \Description{Boxplots of HOF rates vs antenna vendor (top); vs urban/rural areas (bottom).}
     \label{fig:HOF-Vendor-Area-Boxplot}
\end{figure}

Next, we turn our attention to the vendor of the source sector (i.e., antenna vendor). Due to confidentiality issues, we refer to the 4 vendors with the codes V1, V2, V3, and V4, instead of using their actual names. First, we note that different vendors are used in sectors in different regions (North, South, West, \revtwo{Capital area}), Figure \ref{fig:HOF-Vendor-Stats} (top); while all but one vendors are involved in similar proportions in intra 4G/5G-NSA HOs and HOs to 3G, Figure \ref{fig:HOF-Vendor-Stats} (bottom). In Figure \ref{fig:HOF-Vendor-Area-Boxplot} (top), we present the boxplots for the effect of the antenna vendor on HOF rates. In this case, we create one plot for each type of RAT and focus on HOF rates $<1\%$ for 4G/5G-NSA, since the values are concentrated in the low-end of the spectrum. ANOVA tests for each HO type and for all HO types concurrently verify this effect is statistically significant but very small ($(F(3, 4911927) = 30524.85, p < .001; \eta^2 = 0.02, 95\% CI [0.02, 1.00])$). Finally, Figure \ref{fig:HOF-Vendor-Area-Boxplot} (bottom) studies the effect of the area type, where this feature takes two values: rural and urban. We observe a small effect of the area type and indeed performing an ANOVA test, we find it statistically significant but small ($F(2, 4664505)=18559.77$, $p\!<\!.001$, $\eta^2 = 7.90\cdot10^{-3}$, $95\% CI [7.76\cdot10^{-3}, 1.00]$), even when we subset per HO type.

\begin{table}[!t]
{\footnotesize{ \centering
     \caption{{\small{Summary Stats of Dataset.}}}
     \vspace{-2mm}
    \begin{tabular}{l c c c c c c}
    \toprule
    Feature & Min & 1st Qu & Median & Mean & 3rd Qu & Max \\    
    \midrule
    Daily HOs & 1 & 76  & 1989 & 6431  & 8591 & 953287  \\ 
    HOF rate & 0.0 & 0.0  & 0.069 & 6.131  & 4.191 & 100.0  \\        
    \bottomrule
    \label{tab:summary-stats}
    \end{tabular}}}
\end{table}

\begin{table}[!t]
{\footnotesize{
\centering
\caption{{\small{Regression Summary: Linear Model w/o 2G HOs.}}}
\vspace{-3mm}
\begin{tabular}{p{2.65cm} p{1.2cm} c c c}
\toprule
Feature & Coeff. & Std Err & t value & Pr($>|t|$) \\    
\midrule
(Intercept) & $-3.64$ & $0.0185$ & $-196$ & $0$ \\   
HO type: 4G/5G-NSA$\rightarrow$3G & $5.23$ & $0.00120$ & $4348$ & $0$ \\ 
Number of daily HOs & $-1.02\cdot10^{-5}$ & $0$ & $-215$ & $0$\\    
Area Type: Rural & $0.416$ & $0.00273$ & $153$ & $0$ \\            
Area Type: Urban & $0.365$ & $0.00259$ & $141$ & $0$ \\
Antenna Vendor: V2 & $0.0241$ & $0.00166$ & $ 14.5$ & $0$ \\
Antenna Vendor: V3 & $1.00$ & $0.0183$ & $54.6$ & $0$ \\         
Antenna Vendor: V4 & $0.227$ & $0.0199$ & $11.4$ & $0$ \\      
Sector Region: North & $-0.107$ & $0.0184$ & $-5.81$  & $6.14\cdot10^{-9}$ \\ 
Sector Region: South & $-0.0527$ & $0.00160$ & $-32.9$ & $0$\\ 
Sector Region: West & $0.577$ & $0.0184$ & $31.5$ & $0$ \\   
\revtwo{District population}
& $-1.52\cdot10^{-7}$ & $0$ & $-54.7$ & $0$ \\ 
\midrule
\multicolumn{5}{l}{$N=4892154$, \quad RMSE=1.072901, \quad $R^2=0.8502$, \quad AIC=14571839} \\
\bottomrule
\label{tab:regression-features-not2G} 
\end{tabular}
}}
\end{table}

After this first level of analysis, we proceed with regression models that complement those presented in \cref{sec:regression}. Table \ref{tab:regression-linear-all-features} reports the results for a linear regression model, after log-transforming the dependent variable and excluding outliers (i.e., removing entries with HOF rates exceeding 50\%, less than 10 HOs per day or more than 30k HOs per day) that includes all main features of the dataset. In line with the simpler univariate model in \cref{sec:regression}, we see that the HO type remains the main contributing factor on HOF, even when accounting for all other covariates. On the other hand, the rest of the features are significant, yet have a much smaller, often negligible, effect. To further delineate the effect of the other covariates, we repeat the analysis after excluding HOs to 2G since they represent only $0.04\%$ of dataset entries and are skewed towards much higher HOFs (see boxplots). The results are summarized in Table \ref{tab:regression-features-not2G} where we see that the HO type (only related to 3G in this case) is pronounced, the rural/urban feature is significant but the two values have a similar effect, as well as a significant and large effect of the vendor and the region (West). We note these latter findings (effect of Vendor V2 and West) remain significant even if we exclude the HOs to 2G and 3G, and regress only over the intra 4G/5G-NSA HOs.

\begin{table}[!t]
{\footnotesize{
\centering
\caption{{\small{Quantile Regression w/o Outliers.}}}
\vspace{-3mm}
\begin{tabular}{l c c c c}
\toprule
Feature; Quantile & Coeff. & Std Err & t value & Pr($>|t|$) \\    
\midrule
(Intercept); $\tau=0.2$ & $-3.59$ & $0.00072$ & $-5000.50$ & $0$ \\   
HO type: 4G/5G-NSA$\rightarrow$2G & $5.80$ & $0.07401$ & $78.37$ & $0$ \\ 
HO type: 4G/5G-NSA$\rightarrow$3G & $4.86$ & $0.00113$ & $4297.03$ & $0$ \\ 
\midrule
(Intercept); $\tau=0.4$ & $-2.99$ & $0.00077$ & $-3865.27$ & $0$ \\   
HO type: 4G/5G-NSA$\rightarrow$2G & $5.880$ & $0.07951$ & $73.95$ & $0$ \\ 
HO type: 4G/5G-NSA$\rightarrow$3G & $4.79$ & $0.00122$ & $3935.15$ & $0$ \\ 
\midrule
(Intercept); $\tau=0.6$ & $-2.56$ & $0.00066$ & $-3874.20$ & $0$ \\   
HO type: 4G/5G-NSA$\rightarrow$2G & $5.84$ & $0.06822$ & $85.74$ & $0$ \\ 
HO type: 4G/5G-NSA$\rightarrow$3G & $4.83$ & $0.00104$ & $4632.57$ & $0$ \\ 
\midrule
(Intercept) $\tau=0.8$ & $-2.09$ & $0.00092$ & $-2281.89$ & $0$ \\   
HO type: 4G/5G-NSA$\rightarrow$2G & $5.72$ & $0.09450$ & $60.57$ & $0$ \\ 
HO type: 4G/5G-NSA$\rightarrow$3G & $4.97$ & $0.00145$ & $3437.48$ & $0$ \\ 
\bottomrule
\bottomrule 
\end{tabular}\label{tab:regression-quantile}
}}
\end{table}

\balance
As a final robustness test and based on the (near) bimodal distribution of the log-transformed HOF rate variable, we perform quantile regression on 5 intervals ($\tau\in\{0.2,0.4,0.6,0.8\}$), using the HO type as the only feature. Table \ref{tab:regression-quantile} summarizes the results for the case we filter outliers as before, and Table \ref{tab:regression-quantile-All} presents the coefficients for the entire dataset of non-zero HOF rates. These results reinforce the findings of the previous models, verifying the significant and large effect of the HO type on HOFs across the entire spectrum of observed values.

\section{List of Abbreviations}\label{sec:appendix-abbreviations}

To facilitate the reading and understanding of this work, abbreviations and their full terms are provided in Table \ref{tab:abbreviations}.

\begin{table}[!t] 
{\footnotesize{
\centering
\caption{{\small{Quantile Regression -- All HOFs.}}}
\vspace{-3mm}
\begin{tabular}{l c c c c}
\toprule
 & $\tau=0.2$ & $\tau=0.4$ & $\tau=0.6$ & $\tau=0.8$ \\    
\midrule
(Intercept)         & $-3.62$ & $-3.00$ & $-2.58$ & $-2.11$ \\   
HO type: 4G/5G-NSA$\rightarrow$2G & $7.13$ & $7.20$ & $7.13$ & $6.72$ \\ 
HO type: 4G/5G-NSA$\rightarrow$3G & $5.03$ & $4.99$ & $5.15$ & $5.51$ \\ 
\bottomrule
\label{tab:regression-quantile-All} 
\end{tabular}
}}
\end{table}

\begin{table}[H]
{\footnotesize{ 
\centering
\caption{{List of Abbreviations (alphabetically).}}
\begin{tabular}{ll}
\toprule
Abbreviation & Full Term \\    
\midrule
5G-NR & 5G New Radio \\
5G-NSA & 5G-Non-Standalone \\
5G-SA & 5G-Standalone \\
APN & Access Point Name \\
CN & Core Network \\
CS & Circuit Switched \\
DL & Downlink \\
ECDF & Empirical Cumulative Distribution Function \\
EPC & Evolved Packet Core \\
GSM & Global System for Mobile Communications \\
GSMA & Global System for Mobile Communications Association \\
HOF & Handover Failure \\
HO & Handovers \\
IMEI & International Mobile Equipment Identity \\
IMSI & International Mobile Subscriber Identity \\
IoT & Internet-of-Things \\
M2M & Machine-to-Machine \\
MME & Mobile Management Entity \\
MNO & Mobile Network Operator \\
MR & Measurement Report \\
MSC & Mobile Switching Center \\
PS & Packet Switched \\
QoE & Quality of Experience \\
QoS & Quality of Service \\
RACH & Random Access Channel \\
RAN & Radio Access Network \\
RAT & Radio Access Technology \\
RSRQ & Reference Signal Received Quality \\
RRC & Radio Resource Control \\
SGSN & Serving GPRS Support Node \\
SGW & Serving Gateway \\
TAC & Type Allocation Code \\
TAU & Tracking Area Update \\
UE & User Equipment \\
UL & Uplink \\
URLLC & Ultra Reliable Low Latency Communications \\

\midrule
\bottomrule
\label{tab:abbreviations}
\end{tabular}}}
\end{table}

\end{document}